\documentclass[aps,prd,onecolumn,groupedaddress,showpacs,nofootinbib,amssymb]{revtex4}
\usepackage[dvips]{graphicx}
\usepackage{amssymb}
\usepackage{amsmath}
\usepackage{graphicx,color}
\usepackage{amsfonts}
\usepackage{bm}
\usepackage{cancel}
\usepackage{comment}
\usepackage{hyperref}
\usepackage{ulem}

\newcommand{\e}{\mathrm{e}}

\allowdisplaybreaks[4]

\begin{document}

\tolerance=5000

\title{Propagation of Gravitational Waves in a Dynamical Wormhole Background for Two-scalar Einstein--Gauss-Bonnet Theory}

\author{E.~Elizalde$^{1}$}\email{elizalde@ice.csic.es}
\author{Shin'ichi~Nojiri$^{2,3}$}\email{nojiri@gravity.phys.nagoya-u.ac.jp}
\author{S.~D.~Odintsov$^{1,4}$}\email{odintsov@ice.csic.es}
\author{V.~K.~Oikonomou,$^{5}$}\email{voikonomou@gapps.auth.gr; v.k.oikonomou1979@gmail.com}

\affiliation{ $^{1)}$ Institute of Space Sciences (ICE, CSIC) C. Can Magrans s/n, 08193 Barcelona, Spain \\
$^{2)}$ Department of Physics, Nagoya University, Nagoya 464-8602, Japan \\
$^{3)}$ Kobayashi-Maskawa Institute for the Origin of Particles and the Universe, Nagoya University, Nagoya 464-8602, Japan \\
$^{4)}$ ICREA, Passeig Luis Companys, 23, 08010 Barcelona, Spain\\
$^{5)}$ Department of Physics, Aristotle University of
Thessaloniki, Thessaloniki 54124, Greece}

\begin{abstract}
In this work, we propose a model of Einstein--Gauss-Bonnet gravity
coupled with two scalar fields. The two scalar fields are
considered to be ``frozen'' or they become non-dynamical by
employing appropriate constraints in terms of Lagrange multiplier
fields. We show that, even in the case that the arbitrary
spherically symmetric spacetime is dynamical, we can construct a
model where the wormhole spacetime is a stable solution in this
framework. We especially concentrate on the model reproducing the
dynamical wormhole, where the wormhole appears in a finite-time
interval. We investigate the propagation of the gravitational wave
in the wormhole spacetime background and we show that the
propagation speed is different from that of light $\to$ light in
general, and there is a difference in the speeds between the
incoming propagating wave and the outgoing propagating
gravitational wave.
\end{abstract}

\maketitle

\section{Introduction}\label{SecI}

The physical importance of Einstein--Gauss-Bonnet theories is
profound, since such theories emerge as ultraviolet corrections of
the minimal coupled scalar field theory \cite{Codello:2015mba}.
The Einstein--Gauss-Bonnet theory is a viable candidate for
describing the inflationary era \cite{Nojiri:2010wj,
Nojiri:2017ncd, Hwang:2005hb, Nojiri:2006je, Nojiri:2005vv,
Satoh:2007gn, Yi:2018gse, Guo:2009uk, Jiang:2013gza,
Kanti:2015pda, vandeBruck:2017voa, Kanti:1998jd, Pozdeeva:2020apf,
Pozdeeva:2021iwc, Koh:2014bka, Bayarsaikhan:2020jww,
Chervon:2019sey, DeLaurentis:2015fea, Nozari:2017rta,
Odintsov:2018zhw, Kawai:1998ab, Yi:2018dhl, vandeBruck:2016xvt,
Kleihaus:2019rbg, Bakopoulos:2019tvc, Maeda:2011zn,
Bakopoulos:2020dfg, Ai:2020peo, Odintsov:2020xji,
Oikonomou:2020sij, Odintsov:2020zkl, Odintsov:2020mkz,
Venikoudis:2021irr, Kong:2021qiu, Easther:1996yd,
Antoniadis:1993jc, Antoniadis:1990uu, Kanti:1995vq, Kanti:1997br,
Easson:2020mpq, Rashidi:2020wwg, Odintsov:2023aaw,
Odintsov:2023lbb,Oikonomou:2022xoq,Odintsov:2023weg,
Nojiri:2023jtf,TerenteDiaz:2023kgc,Kawai:2023nqs,Kawai:2021edk,Kawai:2017kqt,Choudhury:2023kam},
with the important feature that these models can generate a
desirable blue-tilted tensor spectral index, which is important
since the stochastic gravitational waves produced by an
Einstein--Gauss-Bonnet inflationary era can be detected in the
future gravitational wave experiments. In a recent work
\cite{Nojiri:2020blr}, it was shown that an arbitrary spherically
symmetric spacetime can be realized by using Einstein's gravity
coupled with two scalar fields even if the spacetime is dynamical.
It has also been shown that ghost degrees of freedom inevitably
appear in the model. The ghost has kinetic energy unbounded from
below in classical theory and the ghost generates a negative norm
state in quantum theory as is also well-known in QCD
\cite{Kugo:1979gm}, which conflicts with the Copenhagen
interpretation of quantum theory. Therefore, the existence of the
ghost degrees of freedom conflicts with the physical reality.
Recently, by extending the formulation of
Ref.~\cite{Nojiri:2020blr}, the models realizing a static wormhole
\cite{Nojiri:2023dvf} and gravastar \cite{Nojiri:2023zlp} have
been proposed, see also \cite{Nojiri:2023mbo} for the propagation
of gravity waves in spherically symmetric spacetimes.  The
gravastar is an alternative to the conventional black hole which
might be generated by the Bose-Einstein condensation
\cite{Mazur:2004fk}. In \cite{Nojiri:2020blr} and
\cite{Nojiri:2023dvf}, in order to avoid ghost degrees of freedom,
some constraints on the scalar fields were imposed. The
constraints were similar to the constraint that was used in
mimetic gravity \cite{Chamseddine:2013kea}. Due to the
constraints, the scalar fields become non-dynamical, that is, the
fluctuations of the scalar fields are prohibited and therefore the
scalar fields do not propagate. Although the scalar fields give
effective pressure and effective energy density as in the standard
fluid, the scalar fields do not oscillate. In this sense, the
scalar fields are considered essentially non-dynamical. This
situation is very similar to the original mimetic gravity
approach, where effective dark matter appears. The effective dark
matter has vanishing pressure, as in the standard dust, but the
effective dark matter is non-dynamical and does not collapse by
the gravitational force.

Due to the phenomenological importance of the Einstein--Gauss-Bonnet models, in this paper,
we consider further extension of the model developed in \cite{Nojiri:2020blr} by
introducing a coupling between the Gauss-Bonnet invariant and two scalar fields.
The two scalar fields are considered to be frozen or they become non-dynamical by employing the constraints given by
appropriate Lagrange multiplier fields.
The Gauss-Bonnet term appears as an $\alpha'$ correction from the string theory to
Einstein's gravity \cite{Gross:1986mw}, where $1/\alpha'$ corresponds to the string tension.
We call the model Einstein's gravity including the scalar field(s) and
the Gauss-Bonnet invariant coupled with the scalar fields as the Einstein--Gauss-Bonnet gravity.

In general, however, the propagating speed of the gravitational
wave in the Einstein--Gauss-Bonnet gravity is different from that
of light in the vacuum, (see GW170817 event
works~\cite{TheLIGOScientific:2017qsa, Monitor:2017mdv,
GBM:2017lvd}). In the Friedmann-Lemaitre-Robertson-Walker (FLRW)
Universe background, the function, which describes the coupling
between the scalar field(s) and the Gauss-Bonnet invariant and is
often denoted as $\xi(\phi)$, can be constrained so that the
propagation speed of the gravitational waves to be equal to unity
in natural units, that is, the unit system where the speed of
light in vacuum $c$ is unity, $c=\hbar=1$)~\cite{Odintsov:2020xji,
Odintsov:2020sqy,Oikonomou:2021kql}. It has been, however, shown
in Ref.~\cite{Nojiri:2023jtf} that the propagating speed of the
gravitational wave is always different from that of light in a
spherically symmetric background like black holes or stellar
objects, as long as $\xi(\phi)$ is not a constant, and therefore
such spherically symmetric solutions could not be compatible with
the results of GW170817.

If we consider a model for which the function $\xi(\phi)$ becomes
a constant in the late Universe, or in weak gravity regions, like
for example at the exterior of stellar objects, then no conflict
would exist with the observational data coming from the GW170817,
although the Gauss-Bonnet coupling might play an essential role in
the early Universe or in strong gravity regions. In this paper, we
consider a dynamical wormhole spacetime, which appears only in a
finite time interval in the framework of the
Einstein--Gauss-Bonnet gravity. The function $\xi(\phi)$ becomes
non-trivial only in the region and the time interval that the
wormhole exists. We consider the propagation of the gravitational
wave in the model and we show that the propagation speed of the
gravitational wave could not be equal that of light as expected.
We also find that the speed of the gravitational wave propagating
inwards into the wormhole is different from that of the
gravitational wave exiting from the wormhole.

A wormhole is an exotic compact object, which can be a solution of
the Einstein equations that enable connections between different
regions of the Universe or even between entirely separate
universes~\cite{Misner:1960zz, Wheeler:1957mu}. In order for the
wormhole to exist in the framework of Einstein's gravity, we need
to include matter fluids that violate the null energy condition,
and therefore all the energy conditions. This tells that in
certain reference frames, the energy density of matter can be
perceived as having a negative energy density. In
\cite{Nojiri:2023dvf}, a model whose solution is a wormhole has
been proposed in the framework of Einstein's gravity coupled with
two scalar fields. The energy-momentum tensor of the scalar fields
breaks the energy conditions and one of the scalar fields becomes
a ghost degree of freedom. The ghost can be eliminated by the
Lagrange multiplier constraints and the scalar field is frozen or
becomes non-dynamical. In this work, we shall propose a model of
the Einstein--Gauss-Bonnet gravity coupled with two scalar fields.
By imposing appropriate constraints by using the Lagrange
multiplier fields, the scalar fields are rendered frozen, that is,
the scalar fields become non-dynamical and do not propagate. We
also consider the propagation of gravitational waves in the
context of the Einstein--Gauss-Bonnet gravity coupled with two
scalar fields by also considering the constraints for the scalar
fields. We find the condition that must be satisfied in order for
the propagating speed of the gravitational wave to coincide with
the speed of light. Before considering the model including two
scalar fields, we first consider the model which includes only one
scalar field. In this framework, we review how we can construct a
model that reproduces an arbitrarily given FLRW universe based on
Ref.~\cite{Nojiri:2006je}. We further discuss the propagation of
the gravitational wave during the inflationary era. Accordingly,
we show that, for the arbitrary spherically symmetric spacetime
which even be dynamical, we can construct a model where the
spacetime is a solution in the framework of
two-scalar--Einstein--Gauss-Bonnet gravity, especially
concentrating on the dynamical wormhole, where the wormhole
appears in a finite time interval. We also investigate the
propagation of the gravitational wave in the wormhole background
and show that the propagation speed is different from that of
light in general, and there is a difference in the speeds between
the incoming propagating wave and the outgoing propagating
gravitational wave from the wormhole. We also discuss the energy
conditions corresponding to the wormhole geometry. Furthermore, we
consider other kinds of dynamical spherically symmetric
spacetimes.

\section{Ghost-free Einstein--Gauss-Bonnet Gravity Coupled With Two Scalar Fields}\label{SecII}

In Ref.~\cite{Nojiri:2020blr}, it has been shown that any
spherically symmetric spacetime, even if the spacetime is
dynamical, can be realized by using Einstein's gravity coupled
with two scalar fields. The model in \cite{Nojiri:2020blr},
however, contains ghost degrees of freedom. In this paper, we
first construct models without ghosts by introducing the
constraints given by the Lagrange multiplier fields, which is
similar to the constraint that appeared in the mimetic gravity
theoretical framework~\cite{Chamseddine:2013kea}. Furthermore, in
order to consider the change in the propagating speed of the
gravitational wave, we also include the coupling of the two scalar
fields with the Gauss-Bonnet topological invariant as in
Refs.~\cite{Nojiri:2005vv,Nojiri:2006je} .

We start with the Einstein--Gauss-Bonnet gravity coupled to two
scalar fields $\phi$ and $\chi$, whose action is given by,
\begin{align}
\label{I8}
S_{\phi\chi} = \int d^4 x \sqrt{-g} & \left\{ \frac{R}{2\kappa^2}
 - \frac{1}{2} A (\phi,\chi) \partial_\mu \phi \partial^\mu \phi
 - B (\phi,\chi) \partial_\mu \phi \partial^\mu \chi \right. \nonumber \\
& \left. \qquad - \frac{1}{2} C (\phi,\chi) \partial_\mu \chi \partial^\mu \chi
 - V (\phi,\chi) - \xi(\phi, \chi) \mathcal{G} + \mathcal{L}_\mathrm{matter} \right\}\, .
\end{align}
Here $V(\phi,\chi)$ is the potential for $\phi$ and $\chi$ and $\xi(\phi,\chi)$ is also a function of $\phi$ and $\chi$.
In (\ref{I8}), $\mathcal{G}$ is the Gauss-Bonnet invariant defined by,
\begin{align}
\label{eq:GB}
\mathcal{G} = R^2-4R_{\alpha \beta}R^{\alpha \beta}+R_{\alpha \beta \rho \sigma}R^{\alpha \beta \rho \sigma}\, ,
\end{align}
which is a topological density and a total derivative in four dimensions that gives the Euler number.

By the variation of the action (\ref{I8}) with respect to the
metric $g_{\mu\nu}$, we obtain,
\begin{align}
\label{gb4bD4}
0= &\, \frac{1}{2\kappa^2}\left(- R_{\mu\nu} + \frac{1}{2} g_{\mu\nu} R\right) \nonumber \\
&\, + \frac{1}{2} g_{\mu\nu} \left\{
 - \frac{1}{2} A (\phi,\chi) \partial_\rho \phi \partial^\rho \phi
 - B (\phi,\chi) \partial_\rho \phi \partial^\rho \chi
 - \frac{1}{2} C (\phi,\chi) \partial_\rho \chi \partial^\rho \chi - V (\phi,\chi)\right\} \nonumber \\
&\, + \frac{1}{2} \left\{ A (\phi,\chi) \partial_\mu \phi \partial_\nu \phi
+ B (\phi,\chi) \left( \partial_\mu \phi \partial_\nu \chi
+ \partial_\nu \phi \partial_\mu \chi \right)
+ C (\phi,\chi) \partial_\mu \chi \partial_\nu \chi \right\} \nonumber \\
&\, - 2 \left( \nabla_\mu \nabla_\nu \xi(\phi,\chi)\right)R
+ 2 g_{\mu\nu} \left( \nabla^2 \xi(\phi,\chi)\right)R
+ 4 \left( \nabla_\rho \nabla_\mu \xi(\phi,\chi)\right)R_\nu^{\ \rho}
+ 4 \left( \nabla_\rho \nabla_\nu \xi(\phi,\chi)\right)R_\mu^{\ \rho} \nonumber \\
&\, - 4 \left( \nabla^2 \xi(\phi,\chi) \right)R_{\mu\nu}
 - 4g_{\mu\nu} \left( \nabla_\rho \nabla_\sigma \xi(\phi,\chi) \right) R^{\rho\sigma}
+ 4 \left(\nabla^\rho \nabla^\sigma \xi(\phi,\chi) \right) R_{\mu\rho\nu\sigma}
+ \frac{1}{2} T_{\mathrm{matter}\, \mu\nu} \, ,
\end{align}
and the field equations are obtained by varying the action with
respect to $\phi$ and $\chi$, and these are,
\begin{align}
\label{I10}
0 =& \frac{1}{2} A_\phi \partial_\mu \phi \partial^\mu \phi
+ A \nabla^\mu \partial_\mu \phi + A_\chi \partial_\mu \phi \partial^\mu \chi
+ \left( B_\chi - \frac{1}{2} C_\phi \right)\partial_\mu \chi \partial^\mu \chi
+ B \nabla^\mu \partial_\mu \chi - V_\phi - \xi_\phi \mathcal{G} \, ,\nonumber \\
0 =& \left( - \frac{1}{2} A_\chi + B_\phi \right) \partial_\mu \phi \partial^\mu \phi
+ B \nabla^\mu \partial_\mu \phi
+ \frac{1}{2} C_\chi \partial_\mu \chi \partial^\mu \chi + C \nabla^\mu \partial_\mu \chi
+ C_\phi \partial_\mu \phi \partial^\mu \chi - V_\chi - \xi_\chi \mathcal{G} \, .
\end{align}
Here $A_\phi=\partial A(\phi,\chi)/\partial \phi$, etc. and in
(\ref{gb4bD4}), $T_{\mathrm{matter}\, \mu\nu}$ is the energy-momentum tensor of matter.
We should note that the field equations in (\ref{I10}) are nothing but the Bianchi identities.

We consider a general spherically symmetric and time-dependent
spacetime, whose metric is given by\footnote{ For the argument
that the metric in (\ref{GBiv}) is a general one, see \cite{Nojiri:2020blr}.},
\begin{align}
\label{GBiv}
ds^2 = - \e^{2\nu (t,r)} dt^2 + \e^{2\lambda (t,r)} dr^2 + r^2 \left( d\vartheta^2 + \sin^2\vartheta d\varphi^2 \right)\, .
\end{align}
We also make the following identifications,
\begin{align}
\label{TSBH1}
\phi=t\, , \quad \chi=r\, .
\end{align}
This assumption does not lead to a loss of generality for the following reason:
For the spherically symmetric solutions~(\ref{GBiv}) of the theory~(\ref{I8}), in general,
$\phi$ and $\chi$ depend on both coordinates $t$ and $r$.
Given a solution, the $t$- and $r$-dependence of $\phi$ and $\chi$ can be
determined, and $\phi$ and $\chi$ are then given by specific functions $\phi(t,r)$, $\chi(t,r)$ of $t$ and $r$.
In spacetime regions where these relations are one-to-one, and provided that
$\partial_{\mu} \phi $ is timelike and $\partial_{\mu} \chi$ is spacelike,
one can invert these specific functional forms and
redefine the scalar fields to replace $t$ and $r$ with new scalar
fields, say, $\bar{\phi}$ and $\bar{\chi}$ with $\phi(t,r)\to \phi(\bar{\phi}, \bar{\chi})$
and $\chi(t,r) \to \chi(\bar{\phi}, \bar{\chi})$.
We can then identify the new fields with $t$ and $r$ as in (\ref{TSBH1}),
$\bar{\phi}\to \phi=t$ and $\bar{\chi} \to \chi= r$.
The change of variables $\left( \phi , \chi\right) \rightarrow \left( \bar{\phi} , \bar{\chi} \right)$ can then be
absorbed into the redefinitions of $A$, $B$, $C$, and $V$ in the action~(\ref{I8}).
Therefore, the assumption~(\ref{TSBH1}) does not lead to loss of generality.

As we will see later, in the framework of the two-scalar--Einstein--Gauss-Bonnet theory in (\ref{I8}), we can
construct a model that realizes any given spherically symmetric
geometry in (\ref{GBiv}), which could be static or time-dependent
but $A$ or $C$ often become negative in the realized geometry,
which indicates that $\phi$ or $\chi$ becomes a ghost degree of freedom.
The ghost mode has negative kinetic energy classically and generates negative norm states as a quantum theory.
Therefore, the existence of the ghost mode indicates that the model is physically inconsistent.
We now eliminate ghosts by imposing constraints on $\phi$ and $\chi$, which is similar to the mimetic
constraint in \cite{Chamseddine:2013kea} and make the ghost modes non-dynamical.

For the purpose of eliminating the ghosts by using the Lagrange
multiplier fields $\lambda_\phi$ and $\lambda_\chi$, we add the
following terms to the action $S_{\mathrm{GR} \phi\chi}$ in (\ref{I8}) as
$S_{\mathrm{GR} \phi\chi} \to S_{\mathrm{GR} \phi\chi} + S_\lambda$,
\begin{align}
\label{lambda1}
S_\lambda = \int d^4 x \sqrt{-g} \left[ \lambda_\phi \left( \e^{-2\nu(t=\phi, r=\chi)} \partial_\mu \phi \partial^\mu \phi + 1 \right)
+ \lambda_\chi \left( \e^{-2\lambda(t=\phi, r=\chi)} \partial_\mu \chi \partial^\mu \chi - 1 \right) \right] \, .
\end{align}
The variations of $S_\lambda$ with respect to $\lambda_\phi$ and $\lambda_\chi$ give the constraints,
\begin{align}
\label{lambda2}
0 = \e^{-2\nu(t=\phi, r=\chi)} \partial_\mu \phi \partial^\mu \phi + 1 \, , \quad
0 = \e^{-2\lambda(t=\phi, r=\chi)} \partial_\mu \chi \partial^\mu \chi - 1 \, ,
\end{align}
whose solutions are consistently given by (\ref{TSBH1}).

The constraints in Eq.~(\ref{lambda2}) make the scalar field
$\phi$ and $\chi$ non-dynamical, that is, the fluctuation of
$\phi$ and $\chi$ from the background (\ref{TSBH1}) do not propagate.
In fact, by considering the perturbation from, (\ref{TSBH1}),
\begin{align}
\label{pert1}
\phi=t + \delta \phi \, , \quad \chi=r + \delta \chi\, ,
\end{align}
we obtain by using Eq.~(\ref{lambda2}),
\begin{align}
\label{pert2}
\partial_t \delta \phi = \partial_r \delta \chi = 0\, .
\end{align}
Therefore if we impose the initial condition $\delta\phi=0$, we
find $\delta\phi=0$ in the whole spacetime. On the other hand, if
we impose the boundary condition $\delta\chi\to 0$ when $r\to
\infty$, we find $\delta\chi=0$ in the whole spacetime. These tell
that $\phi$ and $\chi$ are non-dynamical or frozen degrees of
freedom.

We should note that even in the model modified by adding
$S_\lambda$, $\lambda_\phi=\lambda_\chi=0$ is always a solution.
Therefore, any solution for the equations in (\ref{gb4bD4}) and
(\ref{I10}) given by the original action (\ref{I8}) is a solution
of the model whose action is modified by $S_{\mathrm{GR} \phi\chi} \to S_{\mathrm{GR} \phi\chi} + S_\lambda$ in (\ref{lambda1}).
We now validate this statement.

By the modification $S_{\mathrm{GR} \phi\chi} \to S_{\mathrm{GR} \phi\chi} + S_\lambda$ in (\ref{lambda1}),
the equations in (\ref{gb4bD4}) and (\ref{I10}) are modified as follows,
\begin{align}
\label{gb4bD4mod}
0= &\, \frac{1}{2\kappa^2}\left(- R_{\mu\nu} + \frac{1}{2} g_{\mu\nu} R\right) \nonumber \\
&\, + \frac{1}{2} g_{\mu\nu} \left\{
 - \frac{1}{2} A (\phi,\chi) \partial_\rho \phi \partial^\rho \phi
 - B (\phi,\chi) \partial_\rho \phi \partial^\rho \chi
 - \frac{1}{2} C (\phi,\chi) \partial_\rho \chi \partial^\rho \chi - V (\phi,\chi)\right\} \nonumber \\
&\, + \frac{1}{2} \left\{ A (\phi,\chi) \partial_\mu \phi \partial_\nu \phi
+ B (\phi,\chi) \left( \partial_\mu \phi \partial_\nu \chi
+ \partial_\nu \phi \partial_\mu \chi \right)
+ C (\phi,\chi) \partial_\mu \chi \partial_\nu \chi \right\} \nonumber \\
&\, - 2 \left( \nabla_\mu \nabla_\nu \xi(\phi,\chi)\right)R
+ 2 g_{\mu\nu} \left( \nabla^2 \xi(\phi,\chi)\right)R
+ 4 \left( \nabla_\rho \nabla_\mu \xi(\phi,\chi)\right)R_\nu^{\ \rho}
+ 4 \left( \nabla_\rho \nabla_\nu \xi(\phi,\chi)\right)R_\mu^{\ \rho} \nonumber \\
&\, - 4 \left( \nabla^2 \xi(\phi,\chi) \right)R_{\mu\nu}
 - 4g_{\mu\nu} \left( \nabla_\rho \nabla_\sigma \xi(\phi,\chi) \right) R^{\rho\sigma}
+ 4 \left(\nabla^\rho \nabla^\sigma \xi(\phi,\chi) \right) R_{\mu\rho\nu\sigma} \nonumber \\
&\, + \frac{1}{2}g_{\mu\nu} \left\{ \lambda_\phi \left( \e^{-2\nu(r=\chi)} \partial_\rho \phi \partial^\rho \phi + 1 \right)
+ \lambda_\chi \left( \e^{-2\lambda(r=\chi)} \partial_\rho \chi \partial^\rho \chi - 1 \right) \right\} \nonumber \\
&\, - \lambda_\phi \e^{-2\nu(r=\chi)} \partial_\mu \phi \partial_\nu \phi
 - \lambda_\chi \e^{-2\lambda(r=\chi)} \partial_\mu \chi \partial_\nu \chi
+ \frac{1}{2} T_{\mathrm{matter}\, \mu\nu} \, , \\
\label{I10mod}
0 =& \frac{1}{2} A_\phi \partial_\mu \phi \partial^\mu \phi
+ A \nabla^\mu \partial_\mu \phi + A_\chi \partial_\mu \phi \partial^\mu \chi
+ \left( B_\chi - \frac{1}{2} C_\phi \right)\partial_\mu \chi \partial^\mu \chi
+ B \nabla^\mu \partial_\mu \chi - V_\phi - \xi_\phi \mathcal{G} \nonumber \\
&\, - 2 \nabla^\mu \left( \lambda_\phi \e^{-2\nu(r=\chi)} \partial_\mu \phi \right) \, ,\nonumber \\
0 =& \left( - \frac{1}{2} A_\chi + B_\phi \right) \partial_\mu \phi \partial^\mu \phi
+ B \nabla^\mu \partial_\mu \phi
+ \frac{1}{2} C_\chi \partial_\mu \chi \partial^\mu \chi + C \nabla^\mu \partial_\mu \chi
+ C_\phi \partial_\mu \phi \partial^\mu \chi - V_\chi - \xi_\chi \mathcal{G} \nonumber \\
&\, - 2 \nabla^\mu \left( \lambda_\chi \e^{-2\lambda(r=\chi)} \partial_\mu \chi \right)  \, .
\end{align}
We now consider the solution of the equations in (\ref{gb4bD4})
and (\ref{I10}) by assuming (\ref{GBiv}) and (\ref{TSBH1}).
Then, the $(t,t)$ and $(r,r)$ components in (\ref{gb4bD4mod}) give,
\begin{align}
\label{lambdas}
0=\lambda_\phi = \lambda_\chi\, .
\end{align}
Other components in  (\ref{gb4bD4mod}) are satisfied identically.
On the other hand, Eq.~(\ref{I10mod}) gives,
\begin{align}
\label{lambdas2}
0 = \nabla^\mu \left( \lambda_\phi \e^{-2\nu(r=\chi)} \partial_\mu \phi \right)
= \nabla^\mu \left( \lambda_\chi \e^{-2\lambda(r=\chi)} \partial_\mu \chi \right) \, .
\end{align}
Eq.~(\ref{lambdas2}) is satisfied if Eq.~(\ref{lambdas}) is satisfied.
This tells that the solution of Eqs.~(\ref{gb4bD4}) and (\ref{I10})
is a solution of Eqs.~(\ref{gb4bD4mod}) and (\ref{I10mod}) corresponding to the action modified as
$S_{\mathrm{GR} \phi\chi} \to S_{\mathrm{GR} \phi\chi} + S_\lambda$ in (\ref{lambda1})
if $\lambda_\phi$ and $\lambda_\chi$ vanish (\ref{lambdas}).
We should note, however, that for Eqs.~(\ref{gb4bD4mod}) and (\ref{I10mod}),
there could be a solution where $\lambda_\phi$ and $\lambda_\chi$ do not vanish.

\section{Propagation of Gravitational Wave}\label{SecIII}

We now consider the propagation of the gravitational wave.
The condition that the propagating speed of the gravitational wave
in the framework of the Einstein--Gauss-Bonnet gravity coupled
with  one scalar field coincides with the speed of light in the FLRW
metric spacetime has been given in \cite{Oikonomou:2020sij}.
The condition for general spacetime has been given in \cite{Nojiri:2023jtf}.
We reobtain the condition in \cite{Nojiri:2023jtf} even in the case
of the Einstein--Gauss-Bonnet gravity coupled with two scalar fields.

For the following general variation of the metric,
\begin{align}
\label{variation1}
g_{\mu\nu}\to g_{\mu\nu} + h_{\mu\nu}\, ,
\end{align}
we obtain the equation describing the propagation of the gravitational wave as follows,
\begin{align}
\label{gb4bD4B0}
0=&\, \left[ \frac{1}{4\kappa^2} R + \frac{1}{2} \left\{
 - \frac{1}{2} A \partial_\rho \phi \partial^\rho \phi
 - B \partial_\rho \phi \partial^\rho \chi
 - \frac{1}{2} C \partial_\rho \chi \partial^\rho \chi - V \right\}
 - 4 \left( \nabla_{\rho} \nabla_\sigma \xi \right) R^{\rho\sigma} \right] h_{\mu\nu} \nonumber \\
&\, + \bigg[ \frac{1}{4} g_{\mu\nu} \left\{
 - A \partial^\tau \phi \partial^\eta \phi
 - B \left(\partial^\tau \phi \partial^\eta \chi + \partial^\eta \phi \partial^\tau \chi \right)
 - C \partial^\tau \chi \partial^\eta \chi \right\} \nonumber \\
&\, - 2 g_{\mu\nu} \left( \nabla^\tau \nabla^\eta \xi\right)R
 - 4 \left( \nabla^\tau \nabla_\mu \xi\right)R_\nu^{\ \eta} - 4 \left( \nabla^\tau \nabla_\nu \xi\right)R_\mu^{\ \eta}
+ 4 \left( \nabla^\tau \nabla^\eta \xi \right)R_{\mu\nu} \nonumber \\
&\, + 4g_{\mu\nu} \left( \nabla^\tau \nabla_\sigma \xi \right) R^{\eta\sigma}
+ 4g_{\mu\nu} \left( \nabla_\rho \nabla^\tau \xi \right) R^{\rho\eta}
 - 4 \left(\nabla^\tau \nabla^\sigma \xi \right) R_{\mu\ \, \nu\sigma}^{\ \, \eta}
 - 4 \left(\nabla^\rho \nabla^\tau \xi \right) R_{\mu\rho\nu}^{\ \ \ \ \eta}
\bigg] h_{\tau\eta} \nonumber \\
&\, + \frac{1}{2}\left\{ 2 \delta_\mu^{\ \eta} \delta_\nu^{\ \zeta} \left( \nabla_\kappa \xi \right)R
 - 2 g_{\mu\nu} g^{\eta\zeta} \left( \nabla_\kappa \xi \right)R
 - 4 \delta_\rho^{\ \eta} \delta_\mu^{\ \zeta} \left( \nabla_\kappa \xi \right)R_\nu^{\ \rho}
 - 4 \delta_\rho^{\ \eta} \delta_\nu^{\ \zeta} \left( \nabla_\kappa \xi \right)R_\mu^{\ \rho} \right. \nonumber \\
&\, \left. + 4 g^{\eta\zeta} \left( \nabla_\kappa \xi \right) R_{\mu\nu}
+ 4g_{\mu\nu} \delta_\rho^{\ \eta} \delta_\sigma^{\ \zeta} \left( \nabla_\kappa \xi \right) R^{\rho\sigma}
 - 4 g^{\rho\eta} g^{\sigma\zeta} \left( \nabla_\kappa \xi \right) R_{\mu\rho\nu\sigma}
\right\} g^{\kappa\lambda}\left( \nabla_\eta h_{\zeta\lambda} + \nabla_\zeta h_{\eta\lambda} - \nabla_\lambda h_{\eta\zeta} \right) \nonumber \\
&\, + \left\{ \frac{1}{4\kappa^2} g_{\mu\nu} - 2 \left( \nabla_\mu \nabla_\nu \xi\right) + 2 g_{\mu\nu} \left( \nabla^2\xi\right) \right\}
\left\{ -h_{\mu\nu} R^{\mu\nu} + \nabla^\mu \nabla^\nu h_{\mu\nu} - \nabla^2 \left(g^{\mu\nu}h_{\mu\nu}\right) \right\} \nonumber \\
&\, + \frac{1}{2}\left\{ \left( - \frac{1}{2\kappa^2} - 4 \nabla^2 \xi \right) \delta^\tau_{\ \mu} \delta^\eta_{\ \nu}
+ 4 \left( \nabla_\rho \nabla_\mu \xi\right) \delta^\eta_{\ \nu} g^{\rho\tau}
+ 4 \left( \nabla_\rho \nabla_\nu \xi\right) \delta^\tau_{\ \mu} g^{\rho\eta}
 - 4g_{\mu\nu} \nabla^\tau \nabla^\eta \xi \right\} \nonumber \\
&\, \qquad \times \left\{\nabla_\tau\nabla^\phi h_{\eta\phi}
+ \nabla_\eta \nabla^\phi h_{\tau\phi} - \nabla^2 h_{\tau\eta}
 - \nabla_\tau \nabla_\eta \left(g^{\phi\lambda}h_{\phi\lambda}\right)
 - 2R^{\lambda\ \phi}_{\ \eta\ \tau}h_{\lambda\phi}
+ R^\phi_{\ \tau}h_{\phi\eta} + R^\phi_{\ \tau}h_{\phi\eta} \right\} \nonumber \\
&\, + 2 \left(\nabla^\rho \nabla^\sigma \xi \right)
\left\{ \nabla_\nu \nabla_\rho h_{\sigma\mu}
 - \nabla_\nu \nabla_\mu h_{\sigma\rho}
 - \nabla_\sigma \nabla_\rho h_{\nu\mu}
 + \nabla_\sigma \nabla_\mu h_{\nu\rho}
+ h_{\mu\phi} R^\phi_{\ \rho\nu\sigma}
 - h_{\rho\phi} R^\phi_{\ \mu\nu\sigma} \right\} \nonumber \\
&\, + \frac{1}{2}\frac{\partial T_{\mathrm{matter}\,
\mu\nu}}{\partial g_{\tau\eta}}h_{\tau\eta} \, ,
\end{align}
where we have assumed that the matter minimally couples with gravity.
We do not consider the scalar fields $\phi$ and $\chi$,
because the perturbation can be put to vanish as we have observed
in the last section (\ref{pert2}).

We now choose a condition to fix the gauge as follows
\begin{align}
\label{gfc}
0=\nabla^\mu h_{\mu\nu}\, .
\end{align}
Because we are interested in the massless spin-two mode, we also impose the following condition,
\begin{align}
\label{ce}
0=g^{\mu\nu} h_{\mu\nu} \, .
\end{align}
By choosing the conditions in Eqs.~(\ref{gfc}) and (\ref{ce}),
we can reduce Eq.~(\ref{gb4bD4B0}) as follows,
\begin{align}
\label{gb4bD4B}
0=&\, \left[ \frac{1}{4\kappa^2} R + \frac{1}{2} \left\{
 - \frac{1}{2} A \partial_\rho \phi \partial^\rho \phi
 - B \partial_\rho \phi \partial^\rho \chi
 - \frac{1}{2} C \partial_\rho \chi \partial^\rho \chi - V \right\}
 - 4 \left( \nabla_\rho \nabla_\sigma \xi \right) R^{\rho\sigma} \right] h_{\mu\nu} \nonumber \\
&\, + \bigg[ \frac{1}{4} g_{\mu\nu} \left\{
 - A \partial^\tau \phi \partial^\eta \phi
 - B \left(\partial^\tau \phi \partial^\eta \chi + \partial^\eta \phi \partial^\tau \chi \right)
 - C \partial^\tau \chi \partial^\eta \chi \right\} \nonumber \\
&\, - 2 g_{\mu\nu} \left( \nabla^\tau \nabla^\eta \xi\right)R
 - 4 \left( \nabla^\tau \nabla_\mu \xi\right)R_\nu^{\ \eta} - 4 \left( \nabla^\tau \nabla_\nu \xi\right)R_\mu^{\ \eta}
+ 4 \left( \nabla^\tau \nabla^\eta \xi \right)R_{\mu\nu} \nonumber \\
&\, + 4g_{\mu\nu} \left( \nabla^\tau \nabla_\sigma \xi \right) R^{\eta\sigma}
+ 4g_{\mu\nu} \left( \nabla_{\rho} \nabla^\tau \xi \right) R^{\rho\eta}
 - 4 \left(\nabla^\tau \nabla^\sigma \xi \right) R_{\mu\ \, \nu\sigma}^{\ \, \eta}
 - 4 \left(\nabla^\rho \nabla^\tau \xi \right) R_{\mu\rho\nu}^{\ \ \ \ \eta}
\bigg\} h_{\tau\eta} \nonumber \\
&\, + \frac{1}{2}\left\{ 2 \delta_\mu^{\ \eta} \delta_\nu^{\ \zeta} \left( \nabla_\kappa \xi \right)R
 - 4 \delta_\rho^{\ \eta} \delta_\mu^{\ \zeta} \left( \nabla_\kappa \xi \right)R_\nu^{\ \rho}
 - 4 \delta_\rho^{\ \eta} \delta_\nu^{\ \zeta} \left( \nabla_\kappa \xi \right)R_\mu^{\ \rho} \right. \nonumber \\
&\, \left. + 4g_{\mu\nu} \delta_\rho^{\ \eta} \delta_\sigma^{\ \zeta} \left( \nabla_\kappa \xi \right) R^{\rho\sigma}
 - 4 g^{\rho\eta} g^{\sigma\zeta} \left( \nabla_\kappa \xi \right) R_{\mu\rho\nu\sigma}
\right\} g^{\kappa\lambda}\left( \nabla_\eta h_{\zeta\lambda} + \nabla_\zeta h_{\eta\lambda} - \nabla_\lambda h_{\eta\zeta} \right) \nonumber \\
&\, - \left\{ \frac{1}{4\kappa^2} g_{\mu\nu} - 2 \left( \nabla_\mu \nabla_\nu \xi \right) + 2 g_{\mu\nu} \left( \nabla^2\xi \right) \right\}
R^{\mu\nu} h_{\mu\nu} \nonumber \\
&\, + \frac{1}{2}\left\{ \left( - \frac{1}{2\kappa^2} - 4 \nabla^2 \xi \right) \delta^\tau_{\ \mu} \delta^\eta_{\ \nu}
+ 4 \left( \nabla_\rho \nabla_\mu \xi \right) \delta^\eta_{\ \nu} g^{\rho\tau}
+ 4 \left( \nabla_\rho \nabla_\nu \xi \right) \delta^\tau_{\ \mu} g^{\rho\eta}
 - 4g_{\mu\nu} \nabla^\tau \nabla^\eta \xi \right\} \nonumber \\
&\, \qquad \times \left\{ - \nabla^2 h_{\tau\eta} - 2R^{\lambda\ \phi}_{\ \eta\ \tau}h_{\lambda\phi}
+ R^\phi_{\ \tau}h_{\phi\eta} + R^\phi_{\ \tau}h_{\phi\eta} \right\} \nonumber \\
&\, + 2 \left(\nabla^\rho \nabla^\sigma \xi \right)
\left\{ \nabla_\nu \nabla_\rho h_{\sigma\mu}
 - \nabla_\nu \nabla_\mu h_{\sigma\rho}
 - \nabla_\sigma \nabla_\rho h_{\nu\mu}
 + \nabla_\sigma \nabla_\mu h_{\nu\rho}
+ h_{\mu\phi} R^\phi_{\ \rho\nu\sigma}
 - h_{\rho\phi} R^\phi_{\ \mu\nu\sigma} \right\} \nonumber \\
&\, + \frac{1}{2}\frac{\partial T_{\mathrm{matter}\, \mu\nu}}{\partial g_{\tau\eta}}h_{\tau\eta} \, .
\end{align}
In order to investigate the propagating speed $c_\mathrm{GW}$ of the gravitational wave $h_{\mu\nu}$,
we only need to check the parts including the second derivatives of $h_{\mu\nu}$,
\begin{align}
\label{second}
I_{\mu\nu} \equiv&\, I^{(1)}_{\mu\nu} + I^{(2)}_{\mu\nu} \, , \nonumber \\
I^{(1)}_{\mu\nu} \equiv&\, \frac{1}{2}\left\{ \left( - \frac{1}{2\kappa^2} - 4 \nabla^2 \xi \right) \delta^\tau_{\ \mu} \delta^\eta_{\ \nu}
+ 4 \left( \nabla_\rho \nabla_\mu \xi\right) \delta^\eta_{\ \nu} g^{\rho\tau}
+ 4 \left( \nabla_\rho \nabla_\nu \xi\right) \delta^\tau_{\ \mu} g^{\rho\eta}
 - 4g_{\mu\nu} \nabla^\tau \nabla^\eta \xi \right\} \nabla^2 h_{\tau\eta}  \, ,  \nonumber \\
I^{(2)}_{\mu\nu} \equiv &\, 2 \left(\nabla^\rho \nabla^\sigma \xi \right)
\left\{ \nabla_\nu \nabla_\rho h_{\sigma\mu}
 - \nabla_\nu \nabla_\mu h_{\sigma\rho}
 - \nabla_\sigma \nabla_\rho h_{\nu\mu}
 + \nabla_\sigma \nabla_\mu h_{\nu\rho} \right\} \, .
\end{align}
Because we are assuming that the matter minimally couples with
gravity, any contribution of the matter does not couple with any
derivative of $h_{\mu\nu}$ and the contribution does not appear in $I_{\mu\nu}$.
In other words, the matter is not relevant
to the propagating speed of the gravitational wave.
We should note that $I^{(1)}_{\mu\nu}$ does not change the speed of the gravitational
wave from the speed of light.
On the other hand, $I^{(2)}_{\mu\nu}$ changes the speed of the gravitational wave
from that of the light in general.
If $\nabla_\mu \nabla^\nu \xi$ is proportional to the metric $g_{\mu\nu}$,
\begin{align}
\label{condition}
\nabla_\mu \nabla^\nu \xi = \frac{1}{4}g_{\mu\nu} \nabla^2 \xi \, ,
\end{align}
$I^{(2)}_{\mu\nu}$ does not change the speed of the gravitational wave from that of the light.
The condition (\ref{condition}) is identical to that obtained in \cite{Nojiri:2023jtf}
for the Einstein--Gauss-Bonnet gravity coupled with only one scalar field.

Here we have not included the perturbations of the scalar fields
$\phi$ and $\chi$ because the perturbations can be put to vanish
due to the constraints (\ref{lambda2}).
In the case that there are no constraints, or in the case that only one scalar field is
coupled with the Gauss-Bonnet invariant, there are perturbations
of the scalar modes which are mixed with the scalar modes in the
fluctuation of the metric (\ref{variation1}).
As long as we consider the massless spin-two modes, which correspond to the
standard gravitational wave, the spin-two modes decouple
from the scalar mode at the leading order.
If we consider the second-order perturbation, the quadratic moment could appear by the
perturbation of the scalar mode. The quadratic moment plays the
role of the source of the gravitational wave.
Therefore, there could be a difference in the propagation of the gravitational wave
for the cases with and without the constraints (\ref{lambda2}).

\section{Propagation of Gravitational Wave in the FLRW Universe or in the Inflationary Epoch}\label{SecIV}

For comparison, we consider the Einstein--Gauss-Bonnet gravity
coupled with only one scalar field $\phi$, whose action is given by,
\begin{align}
\label{I8one}
S_{\phi\chi} = \int d^4 x \sqrt{-g} & \left\{ \frac{R}{2\kappa^2}
 - \frac{1}{2} \partial_\mu \phi \partial^\mu \phi - V (\phi) - \xi(\phi) \mathcal{G} + \mathcal{L}_\mathrm{matter} \right\}\,  .
\end{align}
Because the scalar field $\phi$ is canonical, we do not impose the
constraints (\ref{lambda2}). Then, the action (\ref{I8one}) can be
regarded as a special case of $A=1$, $B=C=0$, and $V=V(\phi)$ or
$A=B=0$, $C=1$, and $V=V( \phi)$ in the action (\ref{I8}) of the
two scalar fields.

By the variation of the action (\ref{I8one}) with respect to the
metric $g_{\mu\nu}$, we obtain,
\begin{align}
\label{gb4bD4one}
0= &\, \frac{1}{2\kappa^2}\left(- R_{\mu\nu} + \frac{1}{2} g_{\mu\nu} R\right)
+ \frac{1}{2} g_{\mu\nu} \left(
 - \frac{1}{2} \partial_\rho \phi \partial^\rho \phi - V (\phi)\right)
+ \frac{1}{2} \partial_\mu \phi \partial_\nu \phi \nonumber \\
&\, - 2 \left( \nabla_\mu \nabla_\nu \xi(\phi,\chi)\right)R
+ 2 g_{\mu\nu} \left( \nabla^2 \xi(\phi,\chi)\right)R
+ 4 \left( \nabla_\rho \nabla_\mu \xi(\phi,\chi)\right)R_\nu^{\ \rho}
+ 4 \left( \nabla_\rho \nabla_\nu \xi(\phi,\chi)\right)R_\mu^{\ \rho} \nonumber \\
&\, - 4 \left( \nabla^2 \xi(\phi,\chi) \right)R_{\mu\nu}
 - 4g_{\mu\nu} \left( \nabla_\rho \nabla_\sigma \xi(\phi,\chi) \right) R^{\rho\sigma}
+ 4 \left(\nabla^\rho \nabla^\sigma \xi(\phi,\chi) \right) R_{\mu\rho\nu\sigma}
+ \frac{1}{2} T_{\mathrm{matter}\, \mu\nu} \, ,
\end{align}
and the field equation for the scalar field is obtained by varying
the action with respect to $\phi$, and it is given by,
\begin{align}
\label{I10one}
0 =&\, \nabla^\mu \partial_\mu \phi - V' - \xi' \mathcal{G} \, .
\end{align}
In this section, we review the propagation of the gravitational
wave in the FLRW Universe or in the epoch of inflation based on \cite{Nojiri:2023jtf}.

\subsection{Method for Reconstruction}\label{Sec7}

In this subsection, we consider the FLRW Universe with a flat
spatial section, the line element of which is given by,
\begin{align}
\label{FRW}
ds^2= -d{\tilde t}^2 + a(\tilde t)^2\sum_{i=1,2,3} \left(dx^i\right)^2
= -d{\tilde t}^2 + a(\tilde t)^2 \left\{ d{\tilde r}^2 + {\tilde r}^2 \left(d\vartheta^2 + \sin^2\vartheta d\varphi^2 \right) \right\}\, .
\end{align}
Here $a(\tilde t)$ denotes the scale factor and $\tilde t$ is the cosmic time,
which is different from the time coordinate $t$ in (\ref{GBiv}).
Note that the definition of the radial coordinate, $\tilde r$ is also different from $r$ in (\ref{GBiv}).
These coordinates are related by the following equations,
\begin{align}
\label{cv} r= a(\tilde t)\tilde r\, , \quad t=T(\tilde t, \tilde r)
\equiv c_1 \e^{\frac{c_2}{2}{\tilde r}^2 + c_2\int^{\tilde t} \frac{dt'}{a(t') \dot a(t')}}\, ,
\end{align}
where $c_1$ and $c_2$ are constants. In this section, ``$\cdot\ $'' in $\dot a$ means the derivative with respect to $\tilde t$.
Eq.~(\ref{cv}) gives,
\begin{align}
\label{GBiv5}
\e^{2\nu}=&\, \left. \left\{ \left(\frac{\partial T}{\partial \tilde t} \right)^2
 - \frac{\dot a (\tilde t) \tilde r}{a (\tilde t)} \frac{\partial T}{\partial \tilde t} \frac{\partial T}{\partial \tilde r} \right\}^{-1} \right|_{\tilde t = \tilde t(t,r), \, \tilde r = \tilde r(t,r),}\, ,
\nonumber \\
\e^{2\lambda} =&\, \left. \frac{1}{a (\tilde t) \dot a (\tilde t) \tilde r} \frac{\partial T}{\partial \tilde t} \frac{\partial T}{\partial \tilde r}
\left\{ \left(\frac{\partial T}{\partial \tilde t} \right)^2
 - \frac{\dot a (\tilde t) \tilde r}{a (\tilde t)} \frac{\partial T}{\partial \tilde t} \frac{\partial T}{\partial \tilde r} \right\}^{-1} \right|_{\tilde t = \tilde t(t,r), \, \tilde r = \tilde r(t,r),} \, .
\end{align}
We should note that the $(t,r)$ component of the metric vanishes under the redefinitions in (\ref{cv}).

Now we have,
\begin{align}
\label{E2}
& \Gamma^t_{ij}= a^2 H \delta_{ij}\, ,\quad \Gamma^i_{jt}=\Gamma^i_{tj}=H\delta^i_{\ j}\, , \nonumber \\
& R_{itjt}= -\left(\dot H + H^2\right)a^2h_{ij}\, ,\quad
R_{ijkl}= a^4 H^2 \left(\delta_{ik} \delta_{lj} - \delta_{il} \delta_{kj}\right)\, ,\nonumber \\
& R_{tt}=-3\left(\dot H + H^2\right)\, ,\quad R_{ij}= a^2
\left(\dot H + 3H^2\right) \delta_{ij}\, ,\quad R= 6\dot H + 12
H^2\, , \quad \mbox{other components}=0\, ,
\end{align}
therefore if $\xi$ only depends on the cosmic time $\tilde t$, we obtain,
\begin{align}
\label{xis}
\nabla_{\tilde t} \nabla_{\tilde t} \xi= \ddot \xi \, , \quad
\nabla_i \nabla_j \xi = - a^2 H \delta_{ij} \dot \xi \, , \quad \nabla_{\tilde t} \nabla_i \xi = \nabla_i \nabla_{\tilde t} \xi =0 \, , \quad
\nabla^2 \xi = - \ddot \xi - 3 H \dot \xi \, .
\end{align}
Then Eq.~(\ref{condition}) indicates that,
\begin{align}
\label{FLRWcond}
\ddot \xi = H \dot \xi \, ,
\end{align}
whose solution is,
\begin{align}
\label{FLRWsol}
\dot \xi = \xi_0 a(\tilde t)\, .
\end{align}
We should note, however, that the propagating speed of the
gravitational wave cannot be equal to that of light in the
non-trivial spherically symmetric background, as it was shown in \cite{Nojiri:2023jtf}.
Therefore, we may consider the scenario that $I^{(2)}_{ij}$ can be neglected in the late Universe.
This requires that $\xi$ goes to a constant or vanishes in the late
Universe, which indicates that the Gauss-Bonnet term in the action
(\ref{I8one}) becomes a total derivative and does not give any contribution to the expansion of the Universe,
although the term may be important in the early Universe.
If this scenario is realized, then the theory reduces to a scalar-tensor theory in the late Universe.

In the rest of this subsection, we consider how to construct the model realizing the spacetime with the metric in (\ref{FRW}) for
arbitrary $ a(\tilde t)$ based on \cite{Nojiri:2006je}.

The equations corresponding to the FLRW equations have the following forms, which are given by Eq.~(\ref{gb4bD4one})
or we may choose $A=1$, $B=C=0$, and $V=V(\phi)$ in (\ref{gb4bD4}),
\begin{align}
\label{SEGB3}
0=&\, - \frac{3}{\kappa^2}H^2 + \frac{1}{2}{\dot\phi}^2 + V(\phi) + 24 H^3 \frac{d \xi(\phi(\tilde t))}{d\tilde t}\, ,\nonumber \\
0=&\, \frac{1}{\kappa^2}\left(2\dot H + 3 H^2 \right) + \frac{1}{2}{\dot\phi}^2 - V(\phi)
 - 8H^2 \frac{d^2 \xi(\phi(\tilde t))}{d{\tilde t}^2}
 - 16H \dot H \frac{d\xi(\phi(\tilde t))}{d\tilde t} - 16 H^3 \frac{d \xi(\phi(\tilde t))}{dt}  \, , \nonumber \\
0=&\, \ddot \phi + 3H\dot \phi + V'(\phi) + \xi'(\phi) \mathcal{G}\, .
\end{align}
We now use the $e$-foldings number $N$ defined by $a=a_0\e^N$ instead of the cosmic time $\tilde t$.
The equations (\ref{SEGB3}) can be integrated by using the $e$-foldings number $N$ as follows,
\begin{align}
\label{SEGB11}
V(\phi(N)) =&\, \frac{3}{\kappa^2}H(N)^2 - \frac{1}{2}H(N)^2 \phi' (N)^2 - 3\e^N H(N)
\int^N \frac{dN_1}{\e^{N_1}} \left(\frac{2}{\kappa^2} H' (N_1) + H(N_1) \phi'(N_1)^2 \right)\, , \\
\label{SEGB10}
\xi(\phi(N))=&\, \frac{1}{8}\int^N dN_1 \frac{\e^{N_1}}{H(N_1)^3} \int^{N_1} \frac{dN_2}{\e^{N_2}}
\left(\frac{2}{\kappa^2}H' (N_2) + H(N_2) {\phi'(N_2)}^2 \right)\, .
\end{align}
Eqs.~(\ref{SEGB11}) and (\ref{SEGB10}) tell that by using functions $h(N)$ and $ f(\phi)$, if $\xi(\phi)$ and $V(\phi)$ are given by
\begin{align}
\label{SEGB12}
V(\phi) =&\, \frac{3}{\kappa^2}h \left( f(\phi)\right)^2
 - \frac{h\left( f\left(\phi\right)\right)^2}{2 f'(\phi)^2} - 3h\left( f(\phi)\right) \e^{ f(\phi)}
\int^\phi d\phi_1  f'( \phi_1 ) \e^{- f(\phi_1)} \left(\frac{2}{\kappa^2}h'\left( f(\phi_1)\right)
+ \frac{h'\left( f\left(\phi_1\right)\right)}{ f'(\phi_1 )^2} \right)\, , \\
\label{SEGB13}
\xi(\phi) =&\, \frac{1}{8}\int^\phi d\phi_1
\frac{ f'(\phi_1) \e^{ f(\phi_1)} }{h\left( f\left(\phi_1\right)\right)^3}
\int^{\phi_1} d\phi_2  f'(\phi_2) \e^{- f(\phi_2)} \left(\frac{2}{\kappa^2}h'\left( f(\phi_2)\right)
+ \frac{h\left( f(\phi_2)\right)}{ f'(\phi_2)^2} \right)\, ,
\end{align}
a solution of the equations in (\ref{SEGB3}) is given by,
\begin{align}
\label{SEGB14}
\phi= f^{-1}(N)\quad \left(N= f(\phi)\right)\, ,\quad H = h(N) \, .
\end{align}
Therefore, we obtain a general Einstein--Gauss-Bonnet model
realizing the time-evolution of $H$ given by an arbitrary function
$h(N)$ as in (\ref{SEGB14}) \cite{Nojiri:2006je}.

We should note that the history of the expansion of the Universe
is determined only by the function $h(N)$ and does not depend on
the choice of $f(\phi)$.
By using this indefiniteness in the choice of $f(\phi)$, we consider the possibility that $\xi$ goes
to a constant, or vanishes, in the late Universe.
For example, we may assume,
\begin{align}
\label{xi1} \xi = \xi_0 \left( 1 - \e^{-\xi_1 N} \right)\, ,
\end{align}
where $\xi_0$ and $\xi_1$ are constants and we assume $\xi_1$ is positive.
Then $\xi$ rapidly goes to a constant $\xi\to \xi_0$ when $N$ becomes large.
And therefore, the Einstein--Gauss-Bonnet gravity transits to the standard scalar-tensor theory.
The cosmic time of the transition can be adjusted by fine-tuning the parameter $\xi_1$.
For example, if we choose $1/\xi_1 \lesssim 60$, $\xi(\phi)$ becomes almost constant in the epoch of the reheating.

\subsection{Gravitational Waves During the Inflationary Era}

Under the assumption (\ref{xi1}), we may estimate the propagating
speed $c_\mathrm{GW}$ of the gravitational waves.
We now consider the plain wave, $h_{ij} \propto \mathrm{Re} \left( \e^{-i \omega t + i \bm{k}\cdot \bm{x}} \right)$.
Under the condition (\ref{gfc}) and (\ref{ce}), by using (\ref{xis}), Eq.~(\ref{second}) gives
the following dispersion relation for high energy gravitational waves,
\begin{align}
\label{GWSFLRW1} 0 =  \frac{1}{2} \left( - \frac{1}{2\kappa^2} + 4
\left( \ddot \xi + 3 H \dot \xi\right) \right) \left( \omega^2 - \frac{k^2}{a^2} \right) + 2 \ddot \xi \omega^2 \, ,
\end{align}
where $k^2 = \bm{k}\cdot \bm{k}$. Eq.~(\ref{GWSFLRW1}) tells that
\begin{align}
\label{GWSFLRW2}
{c_\mathrm{GW}}^2 = \frac{1 - 8 \kappa^2 \left( \ddot \xi + 3 H \dot \xi\right)}{1 - 8\kappa^2 \left( 2 \ddot \xi
+ 3 H \dot \xi\right)} c^2 \sim \left( 1 + 8 \kappa^2 \ddot \xi \right) c^2 \, .
\end{align}
Now the speed of light $c$ is given by $c^2 = \frac{1}{a^2}$. We
also we assumed $\left| \kappa^2 \ddot \xi  \right| \ll 1$.
Eq.~(\ref{GWSFLRW2}) tells that if $\ddot\xi>0$ $\left(\ddot
\xi<0\right)$, the propagating speed of the gravitational wave is
larger (smaller) than that of light. The event GW170817 indicated
that the propagating speed $c_\mathrm{GW}$ of the gravitational
wave is constrained as follows,
\begin{align}
\label{GWp9} \left| \frac{{c_\mathrm{GW}}^2}{c^2} - 1 \right| < 6
\times 10^{-15}\,  .
\end{align}
Eq.~(\ref{GWSFLRW2}) tells that Eq.~(\ref{GWp9}) gives the following constraint,
\begin{align}
\label{GWp9FLRW}
\left| 8 \kappa^2 \ddot \xi \right| < 6 \times 10^{-15}\, .
\end{align}
Especially, in the case of (\ref{xi1}), we obtain,
\begin{align}
\label{GWp9FLRW}
\left| 8 \kappa^2 \xi_0 \left( \xi_1 H H' - {\xi_1}^2 H^2 \right) \e^{-\xi_1 N} \right| < 6 \times 10^{-15}\, .
\end{align}
In the present Universe, where the speed of the gravitational wave
was measured, we may assume $N=120\ - \ 140$.

We consider the following model in terms of $e$-foldings number
$N$, which satisfies the above conditions,
\begin{align}
\label{Ex2}
H=h(N) =H_0 \left( 1 + \alpha N^\beta \right)^\gamma \, .
\end{align}
When $N$ is small, we find $H \sim H_0 \left( 1 + \alpha \gamma N^\beta \right)$
and when $N$ is large, $H \sim H_1  \alpha^\gamma N^{\beta\gamma}$.
Therefore, $H_1$ and $\alpha$ should be positive so that $H$ is positive.
We also require $\beta>0$ and $\gamma<0$ so that $H$ is a monotonically decreasing function.

The effective equation of state parameter is given by,
\begin{align}
\label{EC3}
w_\mathrm{eff} = - 1 - \frac{2 \alpha \beta \gamma N^{\beta - 1}}{3 \left( 1 + \alpha N^\beta \right) } \, ,
\end{align}
which goes to $-1$ when $N\to 0$ or $N\to \infty$.
When $w_\mathrm{eff} = - \frac{1}{3}$, we find
\begin{align}
\label{EC4}
 - \alpha \beta \gamma N^{\beta - 1} = 1 + \alpha N^\beta \, .
\end{align}
Let the two solutions of (\ref{EC4}) $N=N_1$ and $N=N_2$ $\left(0<N_1<N_2\right)$.
Then the period where $N<N_1$ corresponds to the inflation in the early Universe and the period
where $N>N_2$ to the accelerating expansion in the present Universe.
Then, the parameters $\alpha$ and $\gamma$ are given in
terms of $\beta$, $N_1$, and $N_2$, as follows,
\begin{align}
\label{EC4B}
\alpha = \frac{ {N_2}^{\beta-1} - {N_1}^{\beta-1}}{\left(N_1 N_2\right)^{\beta-1} \left( N_2 - N_1 \right)} \, , \quad
\gamma= - \frac{{N_2}^\beta - {N_1}^\beta}{\beta \left( {N_2}^{\beta-1} - {N_1}^{\beta-1} \right)}\, .
\end{align}
We should note that $\alpha$ and $\gamma$ are positive, as long as $\beta>1$ as we required.
We now estimate the parameters $\alpha$,
$\beta$, and $\gamma$ in order to obtain realistic models
compatible with the constraint on the gravitational wave speed.
Let the beginning of the inflation correspond to $N=0$.
Then the end of the inflation corresponds to $N=N_1=60-70$ and the recombination
(clear up of the Universe) to $N=120-140$.
The redshift of the recombination is $z=1100$.
Because $1+z = 1/a$, where $a$ is the scale factor, we obtain $N_0-N=\ln \left( 1+ z \right)$,
where $N_0$ is the redshift of the present Universe.
We note $\ln 1,100 \sim 7$.
The redshift which corresponds to the beginning of the accelerating expansion
of the late Universe is approximately $0.4$ and $\ln 1.4 \sim 0.3$.
Therefore, $N_2\sim 2N_1$.
In order to estimate $\alpha$ and $\beta$ in (\ref{EC4B}),
we assume $N_2 = 2 N_1 = \mathcal{O}\left( 10^2 \right)$.
Then we find,
\begin{align}
\label{EC4B2}
\alpha = \frac{1 - 2^{1-\beta}}{2^{-\beta} {N_2}^\beta} \, , \quad
\gamma = - \frac{\left( 1 - 2^{-\beta} \right) N_2}{\beta \left( 1 - 2^{1 - \beta} \right)}\, .
\end{align}
In the early Universe, where $N\to 0$, Eq.~(\ref{Ex2}) has the following form,
\begin{align}
\label{EC4B3}
H \sim H_0 \left( 1 + \alpha \gamma N^\beta \right) \, .
\end{align}
Therefore, $H_0$ corresponds to the scale of inflation and we now
choose $H_0 \sim 10^{14}\, \mathrm{GeV} = 10^{23}\, \mathrm{eV}$.
On the other hand, when $N$ is large $\left( N \sim 10^2 \right)$,
which corresponds to the period of the accelerating expansion of
the present Universe, we find,
\begin{align}
\label{EC4B4}
H=H_0 \alpha^ \gamma N^{\beta \gamma} \, ,
\end{align}
which requires $\alpha^\gamma N^{\beta \gamma} \sim 10^{-56}$
because $H\sim 10^{-33}\, \mathrm{eV}$. Since, $N\sim N_2$, by
using (\ref{EC4B2}), we find,
\begin{align}
\label{EC4B5}
\left( 2^\beta - 2 \right)^{- \frac{\left( 1 - 2^{-\beta} \right) N_2}{\beta \left( 1 - 2^{1 - \beta} \right)}}
\sim 10^{-56} \, .
\end{align}
When $\beta \to 1$, we find $\left( 2^\beta - 2 \right)^{-\frac{\left( 1 - 2^{-\beta} \right) N_2} {\beta
\left( 1 - 2^{1 - \beta} \right)}} \to 0$. On the other hand, when $\beta=2$, we
find, $\left( 2^\beta - 2 \right)^{- \frac{\left( 1 - 2^{-\beta} \right) N_2}{\beta \left( 1 - 2^{1 - \beta} \right)}}
= 2^{- \frac{3}{4}N_2 } \sim 10^{-22}\gg 10^{-56}$.
Therefore, there is a solution $\beta$ for Eq.~(\ref{EC4B5}) when $1<\beta<2$.
Then Eq.~(\ref{EC4B2}) tells $\alpha \sim O\left( 10^{-\left(2-4\right)} \right)$ and
$\gamma \sim \mathcal{O}\left( 10^2 \right)$, and therefore these parameters are neither too
small nor large.

We now consider the propagating speed of the gravitational wave during the inflationary era.
The expression of the propagating speed of the gravitational wave in Eq.~(\ref{GWSFLRW2}) is valid,
and at the beginning of the inflation $N\sim 0$, we find,
\begin{align}
\label{inf1}
8\kappa^2 \ddot\xi = 8\kappa^2 \xi_0 \xi_1 \left( \dot H - \xi H^2 \right) \e^{-\xi_1 N}
\sim - 8\kappa^2 {\xi_0}^2 \xi_1 {H_0}^2 \, .
\end{align}
Here we have assumed that Eqs.~(\ref{xi1}) and (\ref{Ex2}) hold true.
On the other hand, at the end of the inflationary era, we obtain,
\begin{align}
\label{GWSFLRW2end}
{c_\mathrm{GW}}^2 \sim \left( 1 - 8 \kappa^2 \xi_0 {H_0}^2 \left( 1 + \alpha {N_1}^\beta \right)^{2\gamma}
\left( \xi_1 + {\xi_1}^2 \right) \e^{-\xi_1 N_1} \right) c^2 \, .
\end{align}
Here we have used Eqs.~(\ref{Ex2}) and (\ref{EC4}) for $N=N_1$. We
now assume that the speed of the gravitational wave is smaller by
$10\%$ than that of light during inflation. Then
Eq.~(\ref{GWSFLRW2}) tells,
\begin{align}
\label{inf2}
8\kappa^2 {\xi_0}^2 \xi_1 {H_0}^2 \sim \frac{1}{10}\, .
\end{align}
If we assume $\xi_1 = \mathcal{O}(1)$, the factor $\e^{-\xi_1 N} $ in (\ref{GWSFLRW2end}) is very small,
\begin{align}
\label{inf3}
\e^{-\xi_1 N} \sim \e^{-60} \fallingdotseq 8.8\times 10^{-27}\, .
\end{align}
Therefore, we expect that the difference between the speed of the
gravitational wave and that of light can be neglected after the
inflationary era, including the epoch of the reheating.

As we have shown the Einstein--Gauss-Bonnet model coupled with one
scalar field (\ref{I8one}) can be regarded as a special case of
the Einstein--Gauss-Bonnet model coupled with two scalar fields (\ref{I8}).
Furthermore, even if we impose the constraints
(\ref{lambda2}) given by the action (\ref{lambda1}), the
background solution in the model without the constraints
(\ref{lambda2}) is the solution of the model with the constraints (\ref{lambda2}).
Therefore, as long as we consider the
gravitational wave which appears as massless spin-two mode in the
perturbation, the situation in this subsection does not change.
We should note, however, that there are perturbations of scalar mode
in the model without the constraints (\ref{lambda2}), which
includes the model with only one scalar field. The scalar mode is
generated by the perturbation of the scalar field, which may be
mixed with the scalar mode of the metric although only the scalar
mode of the metric does not propagate.
As long as we consider the leading order effects, the scalar mode
decouples from the gravitational wave because the scalar mode has spin zero although
the gravitational wave has spin two.
If we consider the next-to-leading order of the perturbation of the scalar mode,
there will appear quadratic order terms in the perturbation of the scalar mode,
which may coupled with the gravitational wave and may
become a source of the gravitational wave.
Such a fluctuation of the scalar mode could be large in the inflationary epoch and it
may generate the primordial gravitational wave.
On the other hand, in the ghost-free model with the constraints (\ref{lambda2}),
there is no fluctuation in the scalar mode.
Therefore, when we consider the generation of the primordial gravitational wave,
the situation could be different in the cases with and without the constraints (\ref{lambda2}).

\section{Reconstruction for Spherically Symmetric Spacetimes}\label{SecV}

We now consider constructing a model with a solution corresponding
to the geometry (\ref{GBiv}) based on \cite{Nashed:2021cfs, Nojiri:2023qgd}.

For the metric~(\ref{GBiv}), the only non-vanishing connection,
coefficients are
\begin{align}
\label{GBv0}
&\Gamma^t_{tt}=\dot\nu \, , \quad \Gamma^r_{tt} = \e^{-2(\lambda - \nu)}\nu' \, , \quad \Gamma^t_{tr}=\Gamma^t_{rt}=\nu'\, , \quad
\Gamma^t_{rr} = \e^{2\lambda - 2\nu}\dot\lambda \, , \quad \Gamma^r_{tr} = \Gamma^r_{rt} = \dot\lambda \, , \nonumber \\
& \Gamma^r_{rr}=\lambda'\, , \Gamma^i_{jk} = \bar{\Gamma} ^i_{jk}\, ,\quad \Gamma^r_{ij}=-\e^{-2\lambda}r \bar{g}_{ij} \, , \quad
\Gamma^i_{rj}=\Gamma^i_{jr}=\frac{1}{r} \, \delta^i_{\ j}\,,
\end{align}
where the metric $\bar{g}_{ij}$ is defined by
$\sum_{i,j=1}^2 \bar{g}_{ij} dx^i dx^j = d\vartheta^2 + \sin^2\vartheta \, d\varphi^2$,
$\left(x^1=\vartheta,\, x^2=\varphi\right)$ and
$\bar{ \Gamma}^i_{jk}$ is the metric connection of $\bar{g}_{ij}$,
while the ``dot'' and the ``prime'' denote differentiation with
respect to $t$ and $r$, respectively.
Using the expression of the Riemann tensor,
\begin{align}
\label{Riemann}
{R^\lambda}_{\ \mu\rho\nu} =\Gamma^\lambda_{\mu\nu,\rho} -\Gamma^\lambda_{\mu\rho,\nu} + \Gamma^\eta_{\mu\nu}\Gamma^\lambda_{\rho\eta}
 - \Gamma^\eta_{\mu\rho}\Gamma^\lambda_{\nu\eta} \,,
\end{align}
one finds,
\begin{align}
\label{curvatures}
R_{rtrt} = & - \e^{2\lambda} \left\{ \ddot\lambda + \left( \dot\lambda - \dot\nu \right) \dot\lambda \right\}
+ \e^{2\nu}\left\{ \nu'' + \left(\nu' - \lambda'\right)\nu' \right\} \, ,\nonumber \\
R_{titj} =& \, r\nu'  \e^{-2\lambda + 2\nu} \bar{g}_{ij} \, ,\nonumber \\
R_{rirj} =& \, \lambda' r \bar{ g}_{ij} \, ,\quad {R_{tirj}= \dot\lambda r \bar{ g}_{ij} } \, , \quad
R_{ijkl} = \left( 1 - \e^{-2\lambda}\right) r^2 \left(\bar{g}_{ik} \bar{g}_{jl} - \bar{g}_{il} \bar{g}_{jk} \right)\, ,\nonumber \\
R_{tt} =& - \left\{ \ddot\lambda + \left( \dot\lambda - \dot\nu \right) \dot\lambda \right\}
+ \e^{-2\lambda + 2\nu} \left\{ \nu'' + \left(\nu' -   \lambda'\right)\nu' + \frac{2\nu'}{r}\right\} \, ,\nonumber \\
R_{rr} =& \, \e^{2\lambda - 2\nu} \left\{ \ddot\lambda + \left( \dot\lambda - \dot\nu \right) \dot\lambda \right\}
 - \left\{ \nu'' + \left(\nu' - \lambda'\right)\nu' \right\}
+ \frac{2 \lambda'}{r} \, ,\quad
R_{tr} =R_{rt} = \frac{2\dot\lambda}{r} \, , \nonumber \\
R_{ij} =&\, \left\{ 1 + \left\{ - 1 - r \left(\nu' - \lambda' \right)\right\} \e^{-2\lambda}\right\} \bar{g}_{ij}\ , \nonumber \\
R=& \, 2 \e^{-2 \nu} \left\{ \ddot\lambda + \left( \dot\lambda -
\dot\nu \right) \dot\lambda \right\} + \e^{-2\lambda}\left\{ -
2\nu'' - 2\left(\nu'  - \lambda'\right)\nu' - \frac{4\left(\nu'  -
\lambda'\right)}{r} + \frac{2\e^{2\lambda} - 2}{r^2} \right\} \, ,
\end{align}
and
\begin{align}
\label{xis}
\nabla_t \nabla_t \xi =&\, \ddot\xi - \dot\nu \dot\xi - \e^{-2(\lambda - \nu)}\nu' \xi' \, , \quad
\nabla_r \nabla_r \xi = \xi'' - \e^{2\lambda - 2\nu} \dot\lambda \dot\xi - \lambda' \xi' \, , \quad
\nabla_t \nabla_r \xi = \nabla_r \nabla_t \xi = {\dot\xi}' - \nu' \dot\xi - \dot\lambda \xi' \, , \nonumber \\
\nabla_i \nabla_j \xi =&\, - \e^{-2\lambda} r {\bar g}_{ij} \xi'
\, , \quad \nabla^2 \xi = - \e^{-2\nu} \left( \ddot\xi - \left(
\dot\nu - \dot\lambda\right) \dot\xi \right) + \e^{-2\lambda}
\left( \xi'' + \left( \nu' - \lambda' - \frac{2}{r} \right) \xi'
\right) \,  .
\end{align}
Then the $(t,t)$, $(r,r)$, $(i,j)$, and $(t,r)$ components of (\ref{gb4bD4}) have the following forms,
\begin{align}
\label{TSBH2aA}
\frac{\e^{-2\lambda + 2\nu}}{\kappa^2} \left( \frac{2\lambda'}{r} + \frac{\e^{2\lambda} - 1}{r^2} \right)
=&\,  - \e^{2\nu} \left( - \frac{A}{2} \e^{-2\nu} - \frac{C}{2} \e^{-2\lambda} - V \right) + J_\mathrm{tmp} \, , \\
\label{TSBH2bA}
\frac{1}{\kappa^2} \left( \frac{2\nu'}{r} - \frac{\e^{2\lambda}  - 1}{r^2} \right)
=&\, \e^{2\lambda} \left( \frac{A}{2} \e^{-2\nu} + \frac{C}{2} \e^{-2\lambda} - V \right) + J_\mathrm{rad} \, , \\
\label{TSBH2cA}
\frac{1}{\kappa^2} \left[ - \e^{-2 \nu} \left\{ \ddot\lambda \right. \right. & + \left. \left. \left( \dot\lambda - \dot\nu \right) \dot\lambda \right\}
+ \e^{-2\lambda}\left( r \left(\nu' - \lambda' \right) + r^2 \nu'' + r^2 \left( \nu' - \lambda' \right) \nu' \right) \right] \nonumber \\
=&\, r^2 \left( \frac{A}{2} \e^{-2\nu} - \frac{C}{2} \e^{-2\lambda} - V \right) + J_\mathrm{ang} \, , \\
\label{TSBH2d} \frac{2\dot\lambda}{\kappa^2 r} =&\, B +
J_\mathrm{mix} \, .
\end{align}
Here
\begin{align}
\label{Js}
J_\mathrm{tmp}\equiv&\,  2 \left\{ \dot\lambda \dot\xi - \e^{-2\lambda+ 2\nu} \left( \xi'' - \left( \lambda' + \frac{2}{r} \right) \xi' \right) \right\} \left[
2 \e^{-2 \nu} \left\{ \ddot\lambda + \left( \dot\lambda - \dot\nu \right) \dot\lambda \right\} \right. \nonumber \\
&\left. + \e^{-2\lambda}\left\{ - 2\nu'' - 2\left(\nu'  - \lambda'\right)\nu' - \frac{4\left(\nu'
 - \lambda'\right)}{r} + \frac{2\e^{2\lambda} - 2}{r^2} \right\} \right] \nonumber \\
&\, - 8 \e^{-2\nu} \left( \ddot\xi - \dot\nu \dot\xi - \e^{-2(\lambda - \nu)}\nu' \xi' \right)
\left[ - \left\{ \ddot\lambda + \left( \dot\lambda - \dot\nu \right) \dot\lambda \right\}
+ \e^{-2\lambda + 2\nu} \left\{ \nu'' + \left(\nu' -   \lambda'\right)\nu' + \frac{2\nu'}{r}\right\} \right] \nonumber \\
&\, - \frac{8}{r^3} \e^{2\nu-2\lambda} \xi' \left[ 1 + \left\{ - 1 - r \left(\nu' - \lambda' \right)\right\} \e^{-2\lambda}\right] \nonumber \\
&\, + 4 \e^{-4\lambda} \left\{  \xi'' - \e^{2\lambda - 2\nu} \dot\lambda \dot\xi - \lambda' \xi' \right\}
\left[ - \e^{2\lambda} \left\{ \ddot\lambda + \left( \dot\lambda - \dot\nu \right) \dot\lambda \right\}
+ \e^{2\nu}\left\{ \nu'' + \left(\nu' - \lambda'\right)\nu' \right\} \right]
 -  \frac{8 \e^{-2\lambda} }{r^2} \xi' \nu' \e^{-2\lambda + 2\nu} \, , \nonumber \\
J_\mathrm{rad}\equiv&\, \left. 2 \left\{  - \e^{2\lambda - 2\nu} \left( \ddot\xi - \dot\nu \dot\xi \right)
+ \left( \nu' - \frac{2}{r} \right) \xi' \right\} \right[
2 \e^{-2 \nu} \left\{ \ddot\lambda + \left( \dot\lambda - \dot\nu \right) \dot\lambda \right\} \nonumber \\
&\, \qquad \qquad \left. + \e^{-2\lambda}\left\{ - 2\nu'' - 2\left(\nu'  - \lambda'\right)\nu' - \frac{4\left(\nu'
 - \lambda'\right)}{r} + \frac{2\e^{2\lambda} - 2}{r^2} \right\} \right] \nonumber \\
&\, + \left[ 8 \e^{-2\lambda} \left\{ \xi'' - \e^{2\lambda - 2\nu} \dot\lambda \dot\xi - \lambda' \xi' \right\} \right]
\left[ \e^{2\lambda - 2\nu} \left\{ \ddot\lambda + \left( \dot\lambda - \dot\nu \right) \dot\lambda \right\}
 - \left\{ \nu'' + \left(\nu' - \lambda'\right)\nu' \right\} + \frac{2 \lambda'}{r} \right] \nonumber \\
&\, + \frac{8}{r^3} \xi' \left[ 1 + \left\{ - 1 - r \left(\nu' - \lambda' \right)\right\} \e^{-2\lambda}\right] \nonumber \\
&\, + 4 \e^{-4\nu} \left\{ \ddot\xi - \dot\nu \dot\xi - \e^{-2(\lambda - \nu)}\nu' \xi' \right\}
\left[ - \e^{2\lambda} \left\{ \ddot\lambda + \left( \dot\lambda - \dot\nu \right) \dot\lambda \right\}
+ \e^{2\nu}\left\{ \nu'' + \left(\nu' - \lambda'\right)\nu' \right\} \right]
 - \frac{8 \e^{-2\lambda}}{r^2} \xi' \lambda' \, , \nonumber \\
J_\mathrm{ang}\equiv&\, 2 r^2\left\{  - \e^{-2\nu} \left( \ddot\xi - \left( \dot\nu - \dot\lambda\right) \dot\xi \right)
+ \e^{-2\lambda} \left( \xi'' + \left( \nu' - \lambda' \right) \xi' \right) \right\} \nonumber \\
&\, \quad \times \left[
2 \e^{-2 \nu} \left\{ \ddot\lambda + \left( \dot\lambda - \dot\nu \right) \dot\lambda \right\}
+ \e^{-2\lambda}\left\{ - 2\nu'' - 2\left(\nu'  - \lambda'\right)\nu' - \frac{4\left(\nu'
 - \lambda'\right)}{r} + \frac{2\e^{2\lambda} - 2}{r^2} \right\} \right] \nonumber \\
&\, + \left[- \frac{8}{r} \e^{-2\lambda} \xi'  - 4 \left\{ - \e^{-2\nu} \left( \ddot\xi - \left( \dot\nu - \dot\lambda\right) \dot\xi \right)
+ \e^{-2\lambda} \left( \xi'' + \left( \nu' - \lambda' - \frac{4}{r} \right) \xi' \right) \right\} \right] \nonumber \\
&\, \quad \times \left[ 1 + \left\{ - 1 - r \left(\nu' - \lambda' \right)\right\} \e^{-2\lambda}\right] \nonumber \\
&\, - 4r^2 \left[ \e^{-4\nu} \left\{ \ddot\xi - \dot\nu \dot\xi - \e^{-2(\lambda - \nu)}\nu' \xi' \right\}
\left[ - \left\{ \ddot\lambda + \left( \dot\lambda - \dot\nu \right) \dot\lambda \right\}
+ \e^{-2\lambda + 2\nu} \left\{ \nu'' + \left(\nu' -   \lambda'\right)\nu' + \frac{\nu'}{r}\right\} \right] \right. \nonumber \\
&\, \left. + \e^{-4\lambda} \left\{ \xi'' - \e^{2\lambda - 2\nu} \dot\lambda \dot\xi - \lambda' \xi' \right\}
\left\{ \e^{2\lambda - 2\nu} \left\{ \ddot\lambda + \left( \dot\lambda - \dot\nu \right) \dot\lambda \right\}
 - \left\{ \nu'' + \left(\nu' - \lambda'\right)\nu' \right\} + \frac{\lambda'}{r} \right\} \right] \nonumber \\
&\, - \frac{4}{r^3} \e^{-2\lambda} \xi' \left[ \left( 1 - \e^{-2\lambda}\right) r^2 \right] \, , \nonumber \\
J_\mathrm{mix}\equiv&\, - 8 \left\{ {\dot\xi}' - \nu' \dot\xi - \dot\lambda \xi' \right\} \left[
 - \e^{-2\nu} \left\{ \ddot\lambda + \left( \dot\lambda - \dot\nu \right) \dot\lambda \right\}
+ \e^{-2\lambda} \left\{ \nu'' + \left(\nu' -   \lambda'\right)\nu' + \frac{\e^{2\lambda} - 1}{2r^2}  \right\} \right]
+ \frac{8}{r^2} \e^{-2\lambda} \xi' \dot\lambda \, .
\end{align}
Eqs.~(\ref{TSBH2aA}), (\ref{TSBH2bA}), (\ref{TSBH2cA}), and
(\ref{TSBH2d}) can be solved as,
\begin{align}
\label{ABCV}
A=& \frac{\e^{2\nu}}{\kappa^2} \left[ - \frac{\e^{-2 \nu}}{r^2} \left\{ \ddot\lambda + \left( \dot\lambda - \dot\nu \right) \dot\lambda \right\}
+ \e^{-2\lambda}\left( \frac{\nu' + \lambda'}{r} + \nu'' +  \left( \nu' - \lambda' \right) \nu' + \frac{\e^{2\lambda} - 1}{r^2}\right) \right]
 - J_\mathrm{tmp} - \frac{\e^{2\nu}}{r^2} J_\mathrm{ang} \, , \nonumber \\
B=&\, \frac{2\dot\lambda}{\kappa^2 r} - J_\mathrm{mix} \, , \nonumber \\
C=&\, \frac{\e^{2\lambda}}{\kappa^2} \left[ \frac{\e^{-2 \nu}}{r^2} \left\{ \ddot\lambda + \left( \dot\lambda - \dot\nu \right) \dot\lambda \right\}
 - \e^{-2\lambda}\left( - \frac{\nu' + \lambda'}{r} + \nu'' +  \left( \nu' - \lambda' \right) \nu' + \frac{\e^{2\lambda} - 1}{r^2}\right) \right]
 - J_\mathrm{tmp} - \frac{\e^{2\nu}}{r^2} J_\mathrm{ang} \, , \nonumber \\
V=& \frac{\e^{-2\lambda}}{\kappa^2} \left( \frac{\lambda' -
\nu'}{r} + \frac{\e^{2\lambda} - 1}{r^2} \right) -
\frac{\e^{-2\nu}}{2} J_\mathrm{tmp} + \frac{\e^{-2\lambda}}{2}
J_\mathrm{rad} \,  .
\end{align}
Then by replacing $(t,r)$ in (\ref{ABCV}) with $(\phi,\chi)$, we
obtain a model which realizes the spacetime given by (\ref{GBiv}).
We should note that there is an ambiguity in the choice of $\xi$.

\subsection{Instant Wormhole Geometry}

In this section, we consider the spacetime where a wormhole appears for a finite time interval and we investigate the gravitational
wave propagating through the wormhole.

The most simple wormhole geometry is given by,
\begin{align}
\label{wh1} \e^{2\nu}=1\, , \quad \e^{2\lambda}=\frac{1}{1 - \frac{r_0}{r}}\, ,
\end{align}
where $r_0$ is the minimum radius of the wormhole.
When $r \sim r_0$, we find $\e^{2\lambda}\sim \frac{r_0}{r - r_0}$, which looks
singular at $r=r_0$.
To avoid the singularity, if $r_0$ is a
constant, by defining a new coordinate $l$, defined as follows,
\begin{align}
\label{lll}
r = \sqrt{ {r_0}^2 + l^2} \, ,\quad dr= \frac{ldl}{\sqrt{ {r_0}^2 + l^2}} \, ,
\end{align}
and then we obtain the following metric,
\begin{align}
\label{lll2}
ds^2 = - dt^2 + \frac{l^2}{{r_0}^2 + l^2 - r_0 \sqrt{ {r_0}^2 + l^2}} dl^2
+ \left(  {r_0}^2 + l^2 \right) \left( d\vartheta^2 + \sin^2\vartheta d\varphi^2 \right)\, . \nonumber \\
\end{align}
When $r\to r_0$, we find $l\to 0$, where $\frac{l^2}{{r_0}^2 + l^2
- r_0 \sqrt{ {r_0}^2 + l^2}} \to 2$, and therefore the spacetime
is regular $r=r_0$, that is, $l=0$.
We analytically continue the value of $l$ in the region where $l$ is negative.
If the region with positive $l$ corresponds to our Universe, the region with
negative $l$ corresponds to another Universe.
The two Universes are connected with each other by the wormhole whose minimum radius
is $r_0$.
We should also note when $\left|l\right|\to \infty$, we
find $\left|l\right|\to r$ and $\frac{l^2}{{r_0}^2 + l^2 - r_0 \sqrt{ {r_0}^2 + l^2}} \to 1$
and therefore the spacetime is asymptotically flat.

We may consider the case that $r_0$ depends on time $r_0=r_0(t)$.
Especially, if $r_0$ is positive only in an interval $-t_0<t<t_0$, the wormhole opens.
For example, we may choose,
\begin{align}
\label{wh2}
r_0 (t) = \frac{l^3 \left( {t_0}^2 - t^2 \right)}{{t_1}^4 + t^4} \, .
\end{align}
Here in addition to $t_0$, $l$ and $t_1$ are constants. We should
note that when $\left| t \right| \to \infty$, the spacetime
becomes flat. In other words, two flat spacetimes are connected by
the wormhole for the finite time interval $-t_0<t<t_0$.

Because $r_0(t)$ is a monotonically decreasing function of $t^2$,
the maximum of the throat radius $r_0(t)$ is given by
$r_0(0)=\frac{l^3{t_0}^2}{{t_1}^4}$.
Let the size of the object traversing the wormhole be $L$.
Then in order for the object to traverse the wormhole, we require,
\begin{align}
\label{tra1}
\frac{l^3{t_0}^2}{{t_1}^4} \gg L \, .
\end{align}
Because the wormhole opens in a finite interval time interval
$-t_0<t<t_0$, we also require that the interval should be large enough.
Let the speed of the object be $V$, the length of the throat is also
given by the order of $r_0(t)$, in order that the object traverses the wormhole, we require,
\begin{align}
\label{tra2}
\frac{r_0(0)}{t_0}=\frac{l^3 t_0}{{t_1}^4} \ll V \leq c \, .
\end{align}
Eqs.~(\ref{tra1}) and (\ref{tra2}) are the necessary conditions so
that the wormhole is traversable.
Furthermore, if the mass of the object is large, the object might break the structure of the
wormhole by the gravitational backreaction.
In our model, the fluctuation in the first-order is prohibited as shown in
(\ref{pert1}) and (\ref{pert2}), and the wormhole could be rather
stable under the backreaction, but if the non-linear effects
become large enough, the higher-order fluctuation may break the
stability of the wormhole.

It might happen that the wormhole in (\ref{wh1}) may connect two
distinct spots in the same Universe.
The distance could be large enough compared with the length of the wormhole $r_0$.
In this case, the wormhole may break causality.
Therefore, if we require causality as a condition for the wormhole, the wormhole in
(\ref{wh1}) could be prohibited.

For the metric (\ref{wh1}), since $\nu$ vanishes, the expressions
in (\ref{xis}) are reduced to,
\begin{align}
\label{xiswh}
\nabla_t \nabla_t \xi =&\, \ddot\xi \, , \quad
\nabla_r \nabla_r \xi = \xi'' - \e^{2\lambda} \dot\lambda \dot\xi - \lambda' \xi' \, , \quad
\nabla_t \nabla_r \xi = \nabla_r \nabla_t \xi = {\dot\xi}' - \dot\lambda \xi' \, , \quad
\nabla_i \nabla_j \xi = - \e^{-2\lambda} r {\bar g}_{ij} \xi' \, , \nonumber \\
\nabla^2 \xi =&\, - \e^{-2\nu} \left( \ddot\xi + \dot\lambda \dot\xi \right)
+ \e^{-2\lambda} \left( \xi'' - \left( \lambda' + \frac{2}{r} \right) \xi' \right) \, ,
\end{align}

We now consider the gravitational wave which propagates in the radial direction,
\begin{align}
\label{hij} h_{ij} = \frac{\mathrm{Re} \left( \e^{-i\omega t + i k r} \right) h_{ij}^{(0)}}{r}
\quad \left( i,j =\theta, \phi, \quad
\sum_i h_{\ \ i}^{(0)\ i}=0 \right)\, , \quad \mbox{other components}=0\, ,
\end{align}
and we also assume $k$ is large enough.
Then Eq.~(\ref{second}) with (\ref{wh1}) and (\ref{xiswh}) gives,
\begin{align}
\label{dis1}
0 =&\, \left[ \left\{ 1 + 8\kappa^2 \left( - \left( 2 \ddot\xi + \dot\lambda \dot\xi \right)
+ \e^{-2\lambda} \left( \xi'' - \left( \lambda' + \frac{2}{r} \right) \xi' \right) \right) \right\}
+ 16 \kappa^2 \e^{-2\lambda} \frac{\xi'}{r} \right] \omega^2 \nonumber \\
&\, - \left[ \left\{  1 + 8 \kappa^2 \left( - \left( \ddot\xi + 2 \dot\lambda \dot\xi \right)
+ \e^{-2\lambda} \left( 2 \xi'' - \left( 2 \lambda' + \frac{2}{r} \right) \xi' \right) \right) \right\}
+ 16 \kappa^2 \e^{-2\lambda} \frac{\xi'}{r} \right] \e^{-2\lambda} k^2 \nonumber \\
&\, + 16 \kappa^2 \e^{-2\lambda} \left(  {\dot\xi}' - \dot\lambda \xi' \right) \omega k \, . \nonumber \\
\end{align}
If $\xi$ is small enough, we find,
\begin{align}
\label{dis2}
\frac{\omega}{k} = \e^{-\lambda} \left[ \pm \left\{ 1 + 4\kappa^2 \left\{ \ddot\xi - \dot\lambda \dot\xi
+ \e^{-2\lambda} \left( \xi'' - \lambda' \xi' \right)
+ 2 \e^{-2\lambda} r \xi' \right\} \right\} - 16 \kappa^2\e^{-\lambda} \left(  {\dot\xi}' - \dot\lambda \xi' \right) \right] \, .
\end{align}
Now the speed of light is given by $c=\e^{-\lambda}$. Therefore,
if $\ddot\xi - \dot\lambda \dot\xi + \e^{-2\lambda} \left( \xi'' -
\lambda' \xi' \right) + 2 \e^{-2\lambda} \frac{\xi'}{r} \mp 4
\e^{-\lambda} \left(  {\dot\xi}' - \dot\lambda \xi' \right)$ is
positive (negative), the propagation speed of the gravitational
wave is larger (smaller) than that of light. In Eq.~(\ref{dis2}),
the $+$ sign corresponds to the gravitational wave outwardly
propagating from the wormhole, and the $-$ sign to that
inward-propagating into the wormhole. Therefore, the speed of
gravitational wave propagating into the wormhole, that is,
\begin{align}
\label{ds3}
v_\mathrm{in} \equiv  \e^{-\lambda} \left[ \left\{ 1 + 4\kappa^2 \left\{ \ddot\xi - \dot\lambda \dot\xi
+ \e^{-2\lambda} \left( \xi'' - \lambda' \xi' \right)
+ 2 \e^{-2\lambda} \frac{\xi'}{r} \right\} \right\} + 16 \kappa^2\e^{-\lambda} \left(  {\dot\xi}' - \dot\lambda \xi' \right) \right] \, ,
\end{align}
is different from the speed of gravitational wave propagating
outwards the wormhole,
\begin{align}
\label{ds4} v_\mathrm{out} \equiv  \e^{-\lambda} \left[ \left\{ 1
+ 4\kappa^2 \left\{ \ddot\xi - \dot\lambda \dot\xi +
\e^{-2\lambda} \left( \xi'' - \lambda' \xi' \right) + 2
\e^{-2\lambda} \frac{\xi'}{r} \right\} \right\} - 16
\kappa^2\e^{-\lambda} \left(  {\dot\xi}' - \dot\lambda \xi'
\right) \right] \, .
\end{align}
This occurs when $\xi$ depends on the radial coordinate $r$ or scalar field $\chi$ and $\lambda$ or $\xi$ depends on the time $t$ or scalar field $\phi$.
If ${\dot\xi}' - \dot\lambda \xi'>0$, we find $v_\mathrm{in}> v_\mathrm{out}$ and if ${\dot\xi}' - \dot\lambda \xi'< 0$, $v_\mathrm{in}< v_\mathrm{out}$.
For example, we may consider the following $\xi(\phi,\chi)$,
\begin{align}
\label{ds4}
\xi(\phi,\chi)= \xi_2 \e^{- \frac{\phi^2}{2{t_2}^2} - \frac{\chi^2}{2{r_1}^2}} = \xi_2 \e^{- \frac{t^2}{2{t_2}^2} - \frac{r^2}{2{r_1}^2}} \, .
\end{align}
Here $\xi_2$ is a constant and $t_2$, and $r_1$ are positive
constants. Under the choice (\ref{ds4}), we find $\xi\to 0$ when
$\left| t \right| \to \infty$ or $r\to \infty$, which tells that
the propagation speed of the gravitational wave coincides with
that of light in the far past or  in the far future or in the
region far from the wormhole. We should also note
$\frac{\dot\xi}{\xi_2}>0$ if $t<0$, $\frac{\xi}{\xi_2}<0$ if
$t>0$, $\frac{\ddot\xi}{\xi_2}>0$ if $\left|t\right|>t_2$, and
$\frac{\ddot\xi}{\xi_2}<0$ if $\left|t\right|<t_2$. We further
find $\frac{\xi'}{\xi_2}<0$, $\frac{\xi''}{\xi_2}>0$ if $r>r_1$,
$\frac{\xi''}{\xi_2}<0$ if $r<r_1$, $\frac{\dot{\xi'}}{\xi_2}<0$
if $t>0$, and $\frac{\dot{\xi'}}{\xi_2}>0$ if $t<0$. On the other
hand, we find $\lambda' = \frac{\frac{r_0}{r^2}}{2\left( 1 -
\frac{r_0}{r} \right)}>0$ and because $\dot\lambda = -
\frac{\frac{\dot r_0}{r}}{2\left( 1 - \frac{r_0}{r} \right)}$, if
we choose $r_0$ by (\ref{wh2}), we obtain $\dot\lambda<0$ if $t<0$
and $\dot\lambda>0$ if $t>0$.

It is complicated and not so useful if we discuss the general
choice of the parameters and general regions, but we may consider
some special simplified cases.
\begin{itemize}
\item First, we consider the region of the throat $r\sim r_0$.
Then in Eqs.~(\ref{ds3}) and (\ref{ds4}), the terms including
$\lambda$ could be dominant because they include the pole
proportional to $\left( r_0 - r \right)^{-1}$ but since
$\e^{-2\lambda} \propto \left( r_0 - r \right)$, the terms
including $\e^{-2\lambda}$ become less dominant.
Therefore, the dominant terms could be given by,
\begin{align}
\label{ds5}
v_\mathrm{in} \sim v_\mathrm{out} \sim c \left[ 1 - 4\kappa^2 \dot\lambda \dot \xi \right]\, .
\end{align}
We should note that the difference between the speed of the
inward-propagating gravitational wave and that of the
outward-propagating one becomes less dominant. If $\xi_2>0$, we
find $\dot\lambda \dot\xi <0$ and the propagating speed of the
gravitational wave becomes larger than that of light. On the other
hand, if $\xi_2<0$, the propagating speed becomes smaller than the
speed of light. \item We also consider the region where $r\gg
r_0,\, r_1$. Then we find,
\begin{align}
\label{ds6}
v_\mathrm{in} \sim v_\mathrm{out} \sim c \left[ 1 + 4\kappa^2 \xi'' \right]\, .
\end{align}
The difference between the speed of the inward-propagating
gravitational wave and that of the outward-propagating one becomes
less dominant, again. If $\xi_2>0$, $\xi''>0$ in the region, and
therefore the propagating speed of the gravitational wave becomes
larger than that of light, again, and if $\xi_2<0$, the
propagating speed becomes smaller than the speed of light.
\end{itemize}
In both of the above regions, the difference between the speed of
the inward-propagating gravitational wave and that of the
outward-propagating one could be neglected. Therefore, a
difference could appear in the intermediate region. We should also
note that in both the above regions, if $\xi_2>0$, the propagating
speed of the gravitational wave becomes larger than that of light
and if $\xi_2<0$, the propagating speed becomes smaller than the
speed of light.

We may speculate regarding the beginning of the Universe.
Through the wormhole, there might be a huge influx of energy from another
spacetime to our spacetime.
The influx may have made a region with very high energy density.
Such a region expands very rapidly by following the first FLRW equation
$\frac{3}{\kappa^2} H^2 = \rho_\mathrm{matter}$,
which may correspond to the generation of the inflationary era.

\subsection{Energy Conditions}

We now study the energy conditions by using the findings of the previous subsection.
The energy conditions in modified gravity
have been well discussed in \cite{Capozziello:2018wul}, especially
for the frame dependence.

There are several kinds of energy conditions. Let $T_{\mu\nu}$ is
the energy-momentum tensor and $\rho$ and $p$ are the energy
density and the pressure, respectively.
\begin{description}
\item[Weak energy condition]
Let $u^\mu$ an arbitrary time-like
vector. The condition is given by $T_{\mu\nu} u^\mu u^\nu \geq 0$.
This means that the energy observed by the observer who is moving
along $u^\mu$ is not negative. In the case of the perfect fluid,
where the pressure $p$ is isotropic, this condition can be written
as $\rho\geq 0$ and $\rho+p\geq 0$.
\item[Dominant energy condition]
Let $u^\mu$ and $v^\mu$ be any pair of causal vector.
Then the conditions are given by $T_{\mu\nu}u^\mu v^\nu \geq 0$
and $\left| T_{\mu\nu} u^\mu v^\nu\right| \leq T_{\mu\nu} u^\mu u^\nu$.
This tells that the energy-momentum flux along $v^\mu$
measured by the observer moving along $u^\mu$ is non-negative.
In the case of the perfect fluid, the conditions can be written as
$\rho \geq 0$ and $\left| \rho + p \right|\leq \rho$.
\item[Strong energy condition]
This condition is given by using an arbitrary time-like vector $u^\mu$ as
$\left( T_{\mu\nu} - \frac{1}{2}g_{\mu\nu} T\right) u^\mu u^\nu \geq 0$.
Here $T$ is the trace of the energy-momentum tensor,
$T\equiv g^{\mu\nu}T_{\mu\nu}$.
In the case of the perfect fluid, this
condition requires $\rho+3p\geq 0$ and $\rho+p \geq 0$.
\item[Null energy condition]
Let $k^\mu$ an arbitrary null vector.
Then the condition is given by $T_{\mu\nu} k^\mu k^\nu \geq 0$.
In the case of the perfect fluid, this condition requires $\rho + p \geq 0$.
\end{description}
Eqs.~(\ref{TSBH2aA}), (\ref{TSBH2bA}), (\ref{TSBH2cA}), and
(\ref{TSBH2d}) indicate that the effective energy density $\rho$,
the effective pressure for radial direction $p_\mathrm{rad}$, the effective pressure
for angular direction $p_\mathrm{ang}$ and the effective energy flow to radial direction $j_r$,
are given by,
\begin{align}
\label{TSBH2aAed}
\rho=&\, \frac{A}{2} \e^{-2\nu} + \frac{C}{2} \e^{-2\lambda} + V + \e^{-2\nu} J_\mathrm{tmp} \, , \\
\label{TSBH2bApr}
p_\mathrm{rad}
=&\, \frac{A}{2} \, \e^{-2\nu} + \frac{C}{2} \, \e^{-2\lambda} - V + \e^{-2\lambda} J_\mathrm{rad} \, , \\
\label{TSBH2cApa}
p_\mathrm{ang}
=&\, \frac{A}{2} \, \e^{-2\nu} - \frac{C}{2} \e^{-2\lambda} - V + \frac{1}{r^2}J_\mathrm{ang} \, , \\
\label{TSBH2djr} j_r =&\, B + J_\mathrm{mix} \,  .
\end{align}
As it is clear from the above expression, the pressure is not
isotropic in general, $p_\mathrm{rad}\neq p_\mathrm{ang}$.
For this case, we need to check the energy condition for each of the pressures.
We should also note that the existence of the energy
flow $j_r$ may make the situation complicated.

For the metric (\ref{wh1}) with (\ref{wh2}), by using
Eqs.~(\ref{TSBH2aA}), (\ref{TSBH2bA}), (\ref{TSBH2cA}), and
(\ref{TSBH2d}), we find,
\begin{align}
\label{TSBH2T2B}
T_{tt} =&\, \rho = 0 \, , \nonumber \\
T_{rr} =&\, \e^{2\lambda} p_\mathrm{rad} = - \e^{2\lambda} \frac{r_0}{\kappa^2 r^3} \, , \nonumber \\
T_{ij} =&\, r^2 {\bar g}_{ij} p_\mathrm{ang} = \frac{1}{\kappa^2} \left[ \frac{\frac{\ddot r_0}{r^4}}{2\left( 1 - \frac{r_0}{r} \right)}
+ \frac{\frac{{\dot r_0}^2}{r^3}}{4\left( 1 - \frac{r_0}{r} \right)^2}
+ \frac{r_0}{2r^3} \right] r^2 \bar{g}_{ij}\, , \nonumber \\
T_{tr} =&\, T_{rt} = \frac{\frac{\dot r_0}{r^2}}{\left( 1 - \frac{r_0}{r} \right)} \, .
\end{align}
Because $\rho=0$, all the energy conditions are violated.
For the usual fluid, this tells that the solution (\ref{wh1}) with
(\ref{wh2}) is totally unstable.
In our model, however, due to the constraints (\ref{lambda2}) given by the action (\ref{lambda2}),
the solution becomes stable. In fact, as discussed in text around
Eq.~(\ref{pert2}), the perturbations of the scalar fields are prohibited.
The only possible fluctuation is the gravitational wave.
Without the constraints (\ref{lambda2}), there is a
perturbation of the scalar fields which are mixed with the scalar
modes in the fluctuation of the metric (\ref{variation1}).
As long as we consider the massless spin-two mode, which corresponds to
the standard gravitational wave, the spin-two mode decouples with
the scalar mode in the leading order.
Although, the second-order perturbation in the perturbation of the scalar modes generates
the quadratic moment, which plays the role of the source of the
gravitational wave.
In the model with the constraints
(\ref{lambda2}), such a mode does not appear and the fluctuation
is only given by the gravitational wave which is always stable.
We should also note that the scalar fields with the constraints
(\ref{lambda2}) cannot be described by perfect fluid or
anisometric compressible fluid.
The scalar fields give some energy
density and pressures but we cannot compress the scalar field
backgrounds in any solution.


\subsection{Embedding of the Static-spherically Symmetric Spacetime in FLRW Spacetime}

In this subsection we consider the embedding of the static and
spherically symmetric spacetime,
\begin{align}
\label{GBivst}
ds^2 = - \e^{2{\tilde \nu} (\tilde r)} d{\tilde t}^2 + \e^{2{\tilde \lambda} (\tilde r)} d{\tilde r}^2
+ {\tilde r}^2 \left( d\vartheta^2 + \sin^2\vartheta d\varphi^2 \right)\, ,
\end{align}
in the FLRW universe by using the conformal time, which we use just for the simplicity of the calculation,
\begin{align}
\label{GBivFLRWconf}
ds^2 = a(\tilde t)^2 \left\{ - \e^{2{\tilde \nu} (\tilde r)} d{\tilde t}^2 + \e^{2{\tilde \lambda} (\tilde r)} d{\tilde r}^2
+ {\tilde r}^2 \left( d\vartheta^2 + \sin^2\vartheta \, d\varphi^2 \right) \right\}\, .
\end{align}
In order to rewrite the above spacetime in the form of (\ref{GBiv}), we define coordinates $(t,r)$ as follows,
\begin{align}
\label{tr1st}
t=\tau \left( \tilde t, \tilde r\right)\, , \quad r=a\left(\tilde t\right) \tilde r \, .
\end{align}
We now find more explicit form of the function $\tau \left( \tilde t, \tilde r\right)$.
Because,
\begin{align}
\label{tr2st}
\left( \begin{array}{c} dt \\ dr \end{array} \right)
= \left( \begin{array}{cc} \frac{\partial\tau}{\partial \tilde t} & \frac{\partial \tau}{\partial \tilde r} \\
\dot a(\tilde t) \tilde r & a(\tilde t) \end{array} \right)
\left( \begin{array}{c} d\tilde t \\ d \tilde r \end{array} \right) \, ,
\end{align}
we obtain,
\begin{align}
\label{tr3st}
\left( \begin{array}{c} d\tilde t \\ d \tilde r \end{array} \right)
= \frac{1}{\frac{\partial\tau}{\partial \tilde t} a(\tilde t) - \frac{\partial \tau}{\partial \tilde r} \dot a(\tilde t) \tilde r}
\left( \begin{array}{cc} a(\tilde t) & - \frac{\partial \tau}{\partial \tilde r} \\
- \dot a(\tilde t) \tilde r & \frac{\partial\tau}{\partial \tilde t} \end{array} \right)
\left( \begin{array}{c} d t \\ d r \end{array} \right) \, ,
\end{align}
and
\begin{align}
\label{tr4st}
ds^2 = \frac{1}{\left( \frac{\partial\tau}{\partial \tilde t} a(\tilde t) - \frac{\partial \tau}{\partial \tilde r} \dot a(\tilde t) \tilde r\right)^2 } &\,
\left\{ - a(\tilde t)^2 \left( a(\tilde t)^2 \e^{2{\tilde \nu} (\tilde r)}
 - {\dot a}(\tilde t)^2 {\tilde r}^2 \e^{2{\tilde \lambda}(\tilde r)} \right) dt^2 \right. \nonumber \\
&\, + a(\tilde t)^2 \left( - \left( \frac{\partial \tau}{\partial \tilde r} \right)^2 \e^{2{\tilde \nu} (\tilde r)}
+ \e^{2{\tilde \lambda} (\tilde r)} \left( \frac{\partial\tau}{\partial \tilde t} \right)^2 \right) dr^2 \nonumber \\
&\, \left. + 2 a(\tilde t)^2 \left( a(\tilde t)  \e^{2{\tilde \nu} (\tilde r)}\frac{\partial \tau}{\partial \tilde r}
 - \e^{2{\tilde \lambda} (\tilde r)} {\dot a}(\tilde t) \tilde r \frac{\partial\tau}{\partial \tilde t} \right) dt dr \right\}
+ r^2 \left( d\vartheta^2 + \sin^2\vartheta d\varphi^2 \right) \, ,
\end{align}
and we require,
\begin{align}
\label{tr5st}
0 = a(\tilde t)  \e^{2{\tilde \nu} (\tilde r)}\frac{\partial \tau}{\partial \tilde r}
 - \e^{2{\tilde \lambda} (\tilde r)} {\dot a}(\tilde t) \tilde r \frac{\partial\tau}{\partial \tilde t} \, .
\end{align}
By assuming $\tau \left( \tilde t, \tilde r\right)=\tau^{(t)} \left( \tilde t\right)\tau^{(r)} \left( \tilde r\right)$, we find
\begin{align}
\label{tr6st} \frac{\e^{2{\tilde \nu} (\tilde r) - 2{\tilde \lambda} (\tilde r)}}{\tilde r \tau^{(r)}
\left( \tilde r\right)}\frac{d \tau^{(r)}\left( \tilde r\right)}{d \tilde r}
= \frac{{\dot a}(\tilde t)}{a(\tilde t) \tau^{(t)} \left( \tilde t\right)} \frac{d\tau^{(t)}\left( \tilde t\right)}{d \tilde t}
= c\, ,
\end{align}
where $c$ is a constant.
The solution of (\ref{tr6st}) is given by,
\begin{align}
\label{tr7st}
\tau^{(t)} \left( \tilde t\right) = \e^{c\int^{\tilde t} dt' \frac{a(t')}{\dot a (t')}}\, , \quad
\tau^{(r)} \left( \tilde r\right) = \e^{c\int^{\tilde r} dr' r' \e^{ - 2{\tilde \nu} (r') + 2{\tilde \lambda} (r')}} \, .
\end{align}
The general solution of $\tau \left( \tilde t, \tilde r\right)$ is given by
\begin{align}
\label{tr7st}
\tau \left( \tilde t, \tilde r\right) = \int dc w(c) \e^{c\left\{ \int^{\tilde t} dt' \frac{a(t')}{\dot a (t')}
+ \int^{\tilde r} dr' r' \e^{ - 2{\tilde \nu} (r') + 2{\tilde \lambda} (r')} \right\}} \, .
\end{align}
Here $w(c)$ is an arbitrary function.
Let assume that Eq.~(\ref{tr1st}) can be solved by using (\ref{tr7st}),
\begin{align}
\label{tr8st}
\tilde t=\tilde t \left( t, r\right)\, , \quad \tilde r= \tilde r \left( t, r \right) \, .
\end{align}
Then by using (\ref{tr4st}), we find,
\begin{align}
\label{tr9st}
\e^{2\nu( t, r )} =&\, \left. \frac{ a(\tilde t)^2 \left( a(\tilde t)^2 \e^{2{\tilde \nu} (\tilde r)} - {\dot a}(\tilde t)^2 {\tilde r}^2 \e^{2{\tilde \lambda}(\tilde r)} \right) }
{\left( \frac{\partial\tau}{\partial \tilde t} a(\tilde t) - \frac{\partial \tau}{\partial \tilde r} \dot a(\tilde t) \tilde r\right)^2 }
\right|_{\tilde t=\tilde t \left( t, r\right), \, \tilde r= \tilde r \left( t, r \right)}
\, , \nonumber \\
\e^{2\lambda(t,r)} =&\, \left. \frac{a(\tilde t)^2 \left( - \left( \frac{\partial \tau}{\partial \tilde r} \right)^2 \e^{2{\tilde \nu} (\tilde r)}
+ \e^{2{\tilde \lambda} (\tilde r)} \left( \frac{\partial\tau}{\partial \tilde t} \right)^2 \right) }
{\left( \frac{\partial\tau}{\partial \tilde t} a(\tilde t) - \frac{\partial \tau}{\partial \tilde r} \dot a(\tilde t) \tilde r\right)^2 }
\right|_{\tilde t=\tilde t \left( t, r\right), \, \tilde r= \tilde r \left( t, r \right)} \, .
\end{align}
Then by using (\ref{ABCV}) and replacing $(t,r)$ in (\ref{ABCV})
with $(\phi,\chi)$, we obtain a model which realizes the spacetime
given by (\ref{GBivFLRWconf}).

We may consider more concrete spacetimes.
If we choose,
\begin{align}
\label{achoice}
a(\tilde t)=\frac{l^n}{{\tilde t}^n}\, ,
\end{align}
$n<0$ corresponds to the decelerating Universe, $n=0$ to the flat spacetime,
$0<n<1$ to the phantom Universe, $n=1$ to the de Sitter Universe,
and $n>1$ to the quintessential Universe.
In the case of (\ref{achoice}), we find
\begin{align}
\label{achoice2}
\int^{\tilde t} dt' \frac{a(t')}{\dot a (t')} = - \frac{1}{2n} {\tilde t}^2 + C_t \, ,
\end{align}
where $C_t$ is a constant of the integration.

In the case of a Schwarzschild spacetime,
\begin{align}
\label{lambdanu}
\e^{2\tilde\nu(\tilde r)}=\e^{-2\lambda(\tilde r)}= 1 - \frac{2M}{r}\, ,
\end{align}
we find,
\begin{align}
\label{lambdanu2}
\int^{\tilde r} dr' r' \e^{ - 2{\tilde \nu} (r') + 2{\tilde \lambda} (r')}
=&\,  \int^{\tilde r} dr' \frac{{r'}^3}{\left( r' - 2M \right)^2} \nonumber \\
=&\, 2M^2 \left\{ \left( \frac{\tilde r}{2M} - 1 \right)^2 + 6
\left( \frac{\tilde r}{2M} - 1 \right) + 6  \left( \frac{\tilde
r}{2M} - 1 \right) - \frac{2}{\frac{\tilde r}{2M} - 1} +C_r
\right\} \, ,
\end{align}
where $C_r$ is a constant of the integration.

In the case of the simplest wormhole,
\begin{align}
\label{swh1}
\e^{2\tilde\nu(\tilde r)}=1\, , \quad \e^{2\lambda(\tilde r)}= \frac{1}{1 - \frac{r_0}{\tilde r}}\, ,
\end{align}
with a constant $r_0$, we find
\begin{align}
\label{swh2}
\int^{\tilde r} dr' r' \e^{ - 2{\tilde \nu} (r') + 2{\tilde \lambda} (r')}
=&\,  \int^{\tilde r} dr' \frac{{r'}^2}{\left( r' - r_0 \right)^2} \nonumber \\
=&\, {r_0}^2 \left\{ \frac{1}{2} \left( \frac{\tilde r}{r_0} - 1
\right)^2 + 2 \left( \frac{\tilde r}{r_0} - 1 \right) + \ln \left(
\frac{\tilde r}{r_0} - 1 \right) + C_r' \right\}\, ,
\end{align}
where $C_r'$ is a constant of the integration.

We now consider the case of a Schwarzschild spacetime
(\ref{lambdanu}) in more detail.
In the standard static case, the Hawking temperature $T_\mathrm{H}$
and the Bekenstein-Hawking entropy $\mathcal{S}$ are given by,
\begin{align}
\label{TS1}
T_\mathrm{H} = \frac{1}{8\pi G M}\, , \quad \mathcal{S} = \frac{A}{4G}\, .
\end{align}
Here $8\pi G = \kappa^2$ and $A$ is the horizon area given by $A=4\pi \left( 2M\right)^2$.
We assume that in the spacetime given
by (\ref{GBivFLRWconf}) with (\ref{lambdanu}), the expansion given
by $a(\tilde t)$ is slow enough.
Then, we may identify $t=a \tilde t$ and $r=a \tilde r$ and in addition,
$\e^{2\tilde \nu (\tilde r )}=\e^{2\nu(r)}$ and
$\e^{2\tilde \\lambda (\tilde r )}=\e^{2\lambda(r)}$, which yield,
\begin{align}
\label{TS2}
\e^{2\nu(r)} = \e^{-\lambda(r)}= 1 - \frac{2aM}{r}\, ,
\end{align}
which relation tells that $M$ scales as $M\to a M$.
Therefore, the Hawking temperature $T_\mathrm{H}$ and the Bekenstein-Hawking
entropy $\mathcal{S}$ are scaled as,
\begin{align}
\label{TS1}
T_\mathrm{H} \to a^{-1} T_\mathrm{H} \, , \quad \mathcal{S} \to a^2 \mathcal{S} \, .
\end{align}
If $M$ corresponds to the thermodynamical energy, the energy
increases by the expansion of the Universe.

In our wormhole model (\ref{wh1}), all the energy conditions are
violated as we have seen in (\ref{TSBH2T2B}) because the energy density vanishes.
This is not so surprising because the energy
density coming from the negative cosmological constant can be
negative in the anti-de Sitter spacetime.
The negative cosmological constant appears as the quantum correction of the vacuum energy coming from fermions.
This may suggest constructing a wormhole solution by using negative vacuum energy as in
\cite{Maldacena:2018lmt}.
As a variation of such a model with
negative energy density, we may consider any relation between the
wormhole and the anti-de Sitter spacetime.

By using the cosmological constant in general $D$-dimensional
spacetime, we write the cosmological constant as,
\begin{align}
\label{E6b} \Lambda=- \frac{(D-1)(D-2)}{l^2}\, ,
\end{align}
where $l$ is the length parameter.
Then we find several
expressions of the anti-de Sitter space in the metric of the warped form in,
\begin{align}
\label{E2b}
ds^2 = \sum_{\mu,\nu=0}^{D-1}=dz^2 + l^2 \e^{2A(z)} \sum_{i,j=0}^{D-2}\tilde g_{ij}(x)
dx^i dx^j \ .
\end{align}
The metric $\tilde g_{ij}$ is the metric of the Einstein manifold which is defined
in terms of the Ricci tensor $\tilde R_{ij}$ constructed from $\tilde g_{ij}$ as follows,
\begin{align}
\label{E3b}
\tilde R_{ij} = k \tilde g_{ij} \, ,
\end{align}
with a constant $k=D-2$, which is equal to zero for the case that
the spacetime given by $\tilde g_{ij}$ is flat, and is equal to
$D-2$ for the case of positively curved spacetime and finally is
equal to $-D+3$ for the case of a negatively curved spacetime.
Depending on the values of $k$, we find several expressions for
$A(z)$ as follows,
\begin{align}
\label{E12b}
A=\left\{\begin{array}{ll}
\frac{z}{l} \quad & (k=0) \\
\ln \cosh \frac{z}{l} \quad & (k=D-2) \\
\ln \sinh \frac{z}{l} \quad & (k=-(D-2)) \\
\end{array}\right. \, .
\end{align}
If we consider the Euclid anti-de Sitter space, the space given by
the metric $\tilde g_{ij}(x)$, we may consider the $D-1$
dimensional sphere for $k=D-2>0$.
Then the radius $l\e^{A(z)}= l \cosh \frac{z}{l}$ of the $D-1$ dimensional sphere has a minimum
at $z=0$, $l\e^{A(0)}=l$ and the radius becomes rapidly larger
when $r$ goes to infinity.
Therefore, this structure is similar to a wormhole.
Instead of the $D-1$ dimensional sphere, we may
consider the $D-1$ dimensional de Sitter space, whose metric we choose,
\begin{align}
\label{dSD-1} {ds_\textrm{$(D-1)$d.dS}}^2 = - d\tau^2 + \cosh^2 \tau {d\Omega_{S_{D-2}}}^2\, ,
\end{align}
where ${d\Omega_{S_{D-2}}}^2$ expresses the line element of $D-2$
dimensional sphere. We obtain the wormhole-like metric as follows,
\begin{align}
\label{mwh}
ds^2 = - l^2 \cosh^2 \frac{z}{l} d\tau^2 + dz^2 + l^2 \cosh^2 \frac{z}{l} \cosh^2 \tau {d\Omega_{S_{D-2}}}^2\, .
\end{align}
The last term could be regarded as the time-dependent wormhole throat.
If we added scalar fields to cancel the cosmological
constant (\ref{E6b}) by the potential for large $z$ as in the
model of this paper, we obtain a model of a wormhole connecting
two flat spacetime.

\section{Summary and Discussion}

In this paper, we proposed a model of Einstein--Gauss-Bonnet
gravity coupled with two scalar fields $\phi$ and $\chi$, whose
action is given in (\ref{I8}). The two scalar fields $\phi$ and
$\chi$ were rendered ``frozen'' or they became non-dynamical by
employing the constraints in (\ref{lambda2}), which are given by
the Lagrange multiplier fields $\lambda_\phi$ and $\lambda_\chi$.
We showed that for an arbitrary spherically symmetric spacetime
which can even be dynamical, as given in (\ref{GBiv}), we can
construct a model where the wormhole spacetime is a solution in
this framework. We especially concentrated on the model
reproducing a dynamical wormhole, where the wormhole appears in a
finite time interval. We investigated the propagation of the
gravitational wave in the wormhole background and showed that the
propagation speed is different from that of light in general, and
there is a difference in the speeds between the incoming in the
wormhole propagating wave and the outgoing propagating one.

We also have to stress that the important point in the present
formulation is that the scalar fields are non-dynamical. The
scalar fields behave as fluids, having energy density and
pressure, but they are not real matter fluids in the following
sense: There is no fluctuation in the scalar fields, therefore not
as in the usual fluid, the sound speed is not generated. When the
energy conditions are violated in the usual fluid, the sound speed
is often larger than the speed of light and it may generate the
violation of causality but such a breakdown does not occur in our
model. The absence of the fluctuation indicates that the solution
is stable under perturbations regardless of the violation of the
energy conditions. The constraints used in this paper mimic the
constraint employed in the framework of mimetic
gravity~\cite{Chamseddine:2013kea}, in which there appears
effective dark matter which behaves as dust matter without
pressure. The effective dark matter is not real matter but
non-dynamical and there does not appear any fluctuations in the
spatial distribution. Furthermore, the effective dark matter does
not collapse by the gravitational force. The formulation in this
paper could be regarded as an extension of the effective dark
matter in the mimetic gravity to a general effective fluid with a
general equation of state. It is also interesting that a dynamical
wormhole of the sort discussed in this work may give the early
Universe epoch alternative to inflation or bounce. Of course, the
detail of such early Universe wormhole era will be discussed
elsewhere.

\section*{Acknowledgements}

This work was partially supported by MICINN (Spain), project
PID2019-104397GB-I00, and by the program Unidad de Excelencia
Maria de Maeztu CEX2020-001058-M, Spain (EE and SDO). The work by
SN was partially supported by the Maria de Maeztu Visiting
Professorship at the Institute of Space Sciences, Barcelona.



\end{document}